
\tolerance=10000
\input phyzzx

\font\mybb=msbm10 at 12pt
\def\bbbb#1{\hbox{\mybb#1}}
\def\Z {\bbbb{Z}}
\def\R {\bbbb{R}}



\def \aa {\alpha}
\def \bb {\beta}
\def \gg {\gamma}

\def \ff {\phi}

\def \ll {\lambda}
\def \mm {\mu}

\def \rr {\rho}
\def \ss {\sigma}
\def \tt {\tau}

\def \www{\Omega}

\def\sym {super Yang-Mills}
\def \ti {\tilde}

\def \2 {{1 \over 2}}
\def \3 {{1 \over 3}}
\def \4 {{1 \over 4}}
\def \5 {{1 \over 5}}
\def \6 {{1 \over 6}}
\def \7 {{1 \over 7}}
\def \8 {{1 \over 8}}
\def \9 {{1 \over 9}}
\def \0 { \infty}

\def\++ {{(+)}}
\def \- {{(-)}}
\def\+-{{(\pm)}}

\def \qq {\qquad}


 \def\unit{\hbox to 3.3pt{\hskip1.3pt \vrule height 7pt width .4pt \hskip.7pt
\vrule height 7.85pt width .4pt \kern-2.4pt
\hrulefill \kern-3pt
\raise 4pt\hbox{\char'40}}}
\def\II{{\unit}}

\def\nup#1({Nucl.\ Phys.\  {\bf B#1}\ (}

\REF\moore{G. Moore, hep-th/9305139,9308052.}
\REF\CJ{E. Cremmer and B. Julia, Phys. Lett. {\bf 80B} (1978) 48; Nucl.
Phys. {\bf B159} (1979) 141.}
\REF\julia{B. Julia in {\it Supergravity and Superspace}, S.W. Hawking and M.
Ro$\check c$ek, C.U.P.
Cambridge,  (1981). }
\REF\julec{B.  Julia, hep-th/9805083.}
\REF\HT{C.M. Hull and P.K. Townsend, hep-th/9410167.}
\REF\gibrap{G.W. Gibbons, hep-th/9803206.}
\REF\buscher{ T. H. Buscher, Phys. Lett. {\bf 159B} (1985) 127, Phys. Lett. {\bf B194}
 (1987), 51 ; Phys. Lett. {\bf B201}
 (1988), 466.}
\REF\rocver {M. Ro\v cek and E. Verlinde, Nucl. Phys.
{\bf B373} (1992), 630.} 
 \REF\givroc {A. Giveon, M.
Ro\v cek, Nucl. Phys. {\bf B380} (1992), 128.}
\REF\alv{E. Alvarez, L. Alvarez-Gaum\' e, J.L. Barbon and Y. Lozano,  Nucl. Phys. {\bf B415} (1994)
71.}
\REF\TD {A. Giveon, M. Porrati and E. Rabinovici, Phys. Rep. {\bf 244}
(1994) 77.}
\REF\HJ{C. M. Hull  and B. Julia, hep-th/9803239.}
\REF\CPS{
E. Cremmer,  I.V. Lavrinenko,  H. Lu,  C.N. Pope,  K.S. Stelle and  T.A. Tran, hep-th/9803259.} 
\REF\Stelle{
K. S. Stelle, hep-th/9803116.}
\REF\ddua {J. Dai, R.G. Leigh and J. Polchinski, Mod. Phys. Lett. {\bf
A4} (1989) 2073.}
\REF\dsei{ M. Dine, P. Huet and N. Seiberg, Nucl. Phys. {\bf B322}
(1989) 301.}
\REF\gibras{  G.W. Gibbons and D.A. Rasheed, hep-th/904177.}
\REF\Bob{B.S. Acharya, M. O'Loughlin and B. Spence, Nucl.
Phys. {\bf B503} (1997) 657; B.S. Acharya, J.M. Figueroa-O'Farrill, M.
O'Loughlin and B. Spence,
hep-th/9707118.}
\REF\thom{M. Blau and G. Thompson, Phys. Lett. {\bf B415} (1997) 242.}
\REF\hawtim{S.W. Hawking,  Phys.Rev. {\bf D46 } (1992) 603.}
\REF\dinst{M.B.  Green, Phys. Lett. {\bf  B354}
(1995) 271,  hep-th/9504108;
M.B.~Green and M.~Gutperle,   Phys.Lett. B398(1997)69, hep-th/9612127;  M.B.~Green and M.~Gutperle,   hep-th/9701093, 
G.~Moore, N.~Nekrasov and  S.~Shatahvilli,   hep-th/9803265; E. Bergshoeff and  K. Behrndt, hep-th/9803090.}
\REF\sevbrane{G.W. Gibbons, M.B.  Green and M.J. Perry, Phys.Lett. B370 (1996) 37, hep-th/9511080.}
\REF\beck{K.~Becker, M.~Becker and  A.~Strominger, hep-th/9507158, Nucl.Phys. {\bf 456} (1995) 130.}
\REF\dinstcal{K.~Becker, M.~Becker, D.R.~Morrison, H.~Ooguri, Y.~Oz and  Z.~Yin,
  hep-th/9608116, Nucl.Phys. {\bf 480} (1996) 225; H.~Ooguri and C.~Vafa,  hep-th/9608079, Phys. Rev. Letts. 77(1996) 3296; M.~Gutperle,   hep-th/9712156.}
\REF\adsstab{P. Breitenlohner and D.Z. Freedman, Phys. Lett. {\bf 115B} (1982) 197; Ann. Phys. {\bf 144} (1982) 197; G.W. Gibbons, C.M. Hull and N.P. Warner, Nucl. Phys. {\bf B218} (1983) 173.}
\REF\mal{J. Maldacena,   hep-th/9711200.}
\REF\OS{K. Osterwalder and R. Schrader, Phys. Rev. Lett.
{\bf 29} (1972) 1423; Helv. Phys. Acta{\bf 46}
 (1973) 277; 
CMP {\bf 31} (1973) 83 and CMP {\bf 42} (1975) 281. 
K. Osterwalder,  in G. Velo and A. Wightman (Eds.)
Constructive Field Theory - Erice lectures 1973, 
Springer-Verlag Berlin 1973; K. Osterwalder in {\it Advances in Dynamical Systems and
Quantum Physics}, Capri conference, World Scientific 1993. }
\REF\vanwick{P. van Nieuwenhuizen and A. Waldron, Phys.Lett. B389 (1996) 29-36, hep-th/9608174. }
\REF\KT{T. Kugo and P.K. Townsend, Nucl. Phys. {\bf B221} (1983) 357.}
\REF\yam{J. Yamron, Phys. Lett. {\bf B213} (1988) 325.}
\REF\vafwit{C. Vafa and E. Witten, Nucl. Phys. {\bf B431} (1994) 3-77,hep-th/9408074.}
\REF\sing{ L. Baulieu, I. Kanno and I. Singer, hep-th/9704167.}
\REF\Seib{N. Seiberg,
hep-th/9705117.}%
\REF\vans{K. Pilch, P. van Nieuwenhuizen and M. Sohnius, Commun. Math. Phys. {\bf 98} (1985) 105.}
\REF\luk{J. Lukierski and A. Nowicki,   Phys.Lett. {\bf 151B}  (1985)  382.}
\REF\witt{E. Witten, hep-th/9802150.}
\REF\bergort{E. Bergshoeff, C.M. Hull and T. Ortin, Nucl. Phys. {\bf B451} (1995) 547, hep-th/9504081.}
\REF\huto{C.M. Hull, Nucl. Phys. {\bf B509} (1998) 252, hep-th/9702067. }
\REF\asp{P. Aspinwall, Nucl. Phys. Proc. Suppl. {\bf  46}  (1996) 30, hep-th/9508154; J. H. Schwarz, hep-th/9508143.}
\REF\fvaf{C. Vafa, Nucl. Phys. {\bf 469} (1996) 403.}
\REF\ythe{C.M. Hull, Nucl.Phys. {\bf B468}  (1996) 113  hep-th/9512181; A. A. Tseytlin Nucl.Phys. 
{\bf B469}  (1996) 51  hep-th/9602064.}
\REF\mythe{I. Bars,  Phys. Rev. {\bf D54} (1996) 5203, hep-th/9604139; hep-th/9604200;
Phys.Rev. {\bf D55} (1997) 2373 hep-th/9607112 .}
\REF\huku{C.M. Hull and R.R. Khuri, in preparation.}
\REF\adssol{M. Gunaydin and N. Marcus, Class. Quant. Grav {\bf 2} (1985) L11; H.J. Kim, L.J. Romans
and P. van Nieuwenhuizen, Phys. Rev. {\bf D32} (1985) 389.}
\REF\gibtow{ G.W. Gibbons and P.K. Townsend, Phys. Rev. Lett. {\bf 71} (1993) 3754. }
\REF\CW{C.M. Hull and N. P. Warner, Class. Quant. Grav. {\bf 5} (1988) 1517.}
\REF\poly{S.S. Gubser,  I. R. Klebanov and  A. M. Polyakov, hep-th/9802109.}
\REF\hor{ G. Horowitz and A. Strominger,
  Nucl. Phys {\bf B360} (1991) 197.}
\REF\huten{C.M. Hull, Phys. Lett. {\bf B357 } (1995) 545, hep-th/9506194.}
\REF\dilres{ G.W. Gibbons, G.T. Horowitz and P.K. Townsend, Class. Quan. Grav. {\bf 12} (1995) 297, hep-th/9410073.}
\REF\gibhaw{G.W. Gibbons and S.W. Hawking, Phys. Rev. {\bf D15} (1977) 2738.}
\REF\kall{P. Claus, R. Kallosh, J. Kumar, P.K. Townsend and A. van Proeyen, hep-th/9801206.}
\REF\witkk{E. Witten, Nucl. Phys. {\bf B195} (1982) 481.}
\REF\haweuc{S.W. Hawking, in {\it General Relativity}, ed. by S.W. Hawking and W. Israel, Cambridge University Press, 1979.}


\Pubnum{ \vbox{  \hbox {QMW-PH-98-28} 
\hbox{hep-th/9806146}} }
\pubtype{}
\date{ June 1998}

\titlepage

\title {\bf    Timelike T-Duality, de Sitter Space, Large $N$ Gauge Theories and Topological Field Theory
 }

\author{C.M. Hull}
\address{Physics Department, Queen Mary and Westfield College,
\break Mile End Road, London E1 4NS, U.K.}
\vskip 0.5cm

\abstract { T-Duality on a timelike circle does not interchange   IIA and IIB string theories, but takes the IIA theory to a
type
$IIB^*$ theory and the IIB theory to a type
$IIA^*$ theory. The type $II^*$ theories admit E-branes, which are the images of the type II D-branes under timelike
T-duality and correspond to imposing Dirichlet boundary conditions in time as well as some of the spatial directions.
The effective action describing an E$n$-brane is the $n$-dimensional Euclidean super-Yang-Mills theory   obtained by
dimensionally reducing  9+1 dimensional super-Yang-Mills on $9-n$ spatial dimensions and one time dimension. The $IIB^*$ theory
has a solution which is the product of 5-dimensional de Sitter space and a 5-hyperboloid, and the E$4$-brane corresponds to a
non-singular  complete solution which interpolates between this solution and flat space. This leads to a duality between the
large $N$ limit of the Euclidean 4-dimensional $U(N)$ super-Yang-Mills theory and the $IIB^*$ string theory in de Sitter
space, and both are invariant under the same de Sitter supergroup. This theory can be twisted to obtain a large $N$
topological gauge theory and its  topological string theory dual. Flat space-time appears to be an unstable vacuum for the 
type $II^*$ theories, but they have  supersymmetric cosmological solutions.}

\endpage

%
%
%
%
\newhelp\stablestylehelp{You must choose a style between 0 and 3.}%
\newhelp\stablelinehelp{You should not use special hrules when stretching
a table.}%
\newhelp\stablesmultiplehelp{You have tried to place an S-Table inside another
S-Table.  I would recommend not going on.}%
%
%
\newdimen\stablesthinline
\stablesthinline=0.4pt
\newdimen\stablesthickline
\stablesthickline=1pt
%
%
\newif\ifstablesborderthin
\stablesborderthinfalse
\newif\ifstablesinternalthin
\stablesinternalthintrue
\newif\ifstablesomit
\newif\ifstablemode
\newif\ifstablesright
\stablesrightfalse
%
%
\newdimen\stablesbaselineskip
\newdimen\stableslineskip
\newdimen\stableslineskiplimit
%
%
\newcount\stablesmode
\newcount\stableslines
\newcount\stablestemp
\stablestemp=3
\newcount\stablescount
\stablescount=0
\newcount\stableslinet
\stableslinet=0
%
%
%
\newcount\stablestyle
\stablestyle=0
%
%
\def\stablesleft{\quad\hfil}%
\def\stablesright{\hfil\quad}%
%
%
\catcode`\|=\active%
%
%
\newcount\stablestrutsize
\newbox\stablestrutbox
\setbox\stablestrutbox=\hbox{\vrule height10pt depth5pt width0pt}
\def\stablestrut{\relax\ifmmode%
                         \copy\stablestrutbox%
                       \else%
                         \unhcopy\stablestrutbox%
                       \fi}%
%
%
\newdimen\stablesborderwidth
\newdimen\stablesinternalwidth
\newdimen\stablesdummy
\newcount\stablesdummyc
\newif\ifstablesin
\stablesinfalse
%
%
\def\begintable{\stablestart%
  \stablemodetrue%
  \stablesadj%
  \halign%
  \stablesdef}%
\def\stablesadj{%
  \ifcase\stablestyle%
    \hbox to \hsize\bgroup\hss\vbox\bgroup%
  \or%
    \hbox to \hsize\bgroup\vbox\bgroup%
  \or%
    \hbox to \hsize\bgroup\hss\vbox\bgroup%
  \or%
    \hbox\bgroup\vbox\bgroup%
  \else%
    \errhelp=\stablestylehelp%
    \errmessage{Invalid style selected, using default}%
    \hbox to \hsize\bgroup\hss\vbox\bgroup%
  \fi}%
\def\stablesend{\egroup%
  \ifcase\stablestyle%
    \hss\egroup%
  \or%
    \hss\egroup%
  \or%
    \egroup%
  \or%
    \egroup%
  \else%
    \hss\egroup%
  \fi}%
\def\stablestart{%
  \ifstablesin%
    \errhelp=\stablesmultiplehelp%
    \errmessage{An S-Table cannot be placed within an S-Table!}%
  \fi
  \global\stablesintrue%
  \global\advance\stablescount by 1%
  \message{<S-Tables Generating Table \number\stablescount}%
  \begingroup%
  \stablestrutsize=\ht\stablestrutbox%
  \advance\stablestrutsize by \dp\stablestrutbox%
  \ifstablesborderthin%
    \stablesborderwidth=\stablesthinline%
  \else%
    \stablesborderwidth=\stablesthickline%
  \fi%
  \ifstablesinternalthin%
    \stablesinternalwidth=\stablesthinline%
  \else%
    \stablesinternalwidth=\stablesthickline%
  \fi%
  \tabskip=0pt%
  \stablesbaselineskip=\baselineskip%
  \stableslineskip=\lineskip%
  \stableslineskiplimit=\lineskiplimit%
  \offinterlineskip%
  \def\borderrule{\vrule width \stablesborderwidth}%
  \def\internalrule{\vrule width \stablesinternalwidth}%
  \def\thinline{\noalign{\hrule height \stablesthinline}}%
  \def\thickline{\noalign{\hrule height \stablesthickline}}%
  \def\trule{\omit\leaders\hrule height \stablesthinline\hfill}%
  \def\ttrule{\omit\leaders\hrule height \stablesthickline\hfill}%
  \def\tttrule##1{\omit\leaders\hrule height ##1\hfill}%
  \def\stablesel{&\omit\global\stablesmode=0%
    \global\advance\stableslines by 1\borderrule\hfil\cr}%
  \def\el{\stablesel&}%
  \def\elt{\stablesel\thinline&}%
  \def\eltt{\stablesel\thickline&}%
  \def\elttt##1{\stablesel\noalign{\hrule height ##1}&}%
  \def\elspec{&\omit\hfil\borderrule\cr\omit\borderrule&%
              \ifstablemode%
              \else%
                \errhelp=\stablelinehelp%
                \errmessage{Special ruling will not display properly}%
              \fi}%
  \def\stmultispan##1{\mscount=##1 \loop\ifnum\mscount>3 \stspan\repeat}%
  \def\stspan{\span\omit \advance\mscount by -1}%
  \def\multicolumn##1{\omit\multiply\stablestemp by ##1%
     \stmultispan{\stablestemp}%
     \advance\stablesmode by ##1%
     \advance\stablesmode by -1%
     \stablestemp=3}%
  \def\multirow##1{\stablesdummyc=##1\parindent=0pt\setbox0\hbox\bgroup%
    \aftergroup\emultirow\let\temp=}
  \def\emultirow{\setbox1\vbox to\stablesdummyc\stablestrutsize%
    {\hsize\wd0\vfil\box0\vfil}%
    \ht1=\ht\stablestrutbox%
    \dp1=\dp\stablestrutbox%
    \box1}%
  \def\stpar##1{\vtop\bgroup\hsize ##1%
     \baselineskip=\stablesbaselineskip%
     \lineskip=\stableslineskip%
     \lineskiplimit=\stableslineskiplimit\bgroup\aftergroup\estpar\let\temp=}%
  \def\estpar{\vskip 6pt\egroup}%
  \def\stparrow##1##2{\stablesdummy=##2%
     \setbox0=\vtop to ##1\stablestrutsize\bgroup%
     \hsize\stablesdummy%
     \baselineskip=\stablesbaselineskip%
     \lineskip=\stableslineskip%
     \lineskiplimit=\stableslineskiplimit%
     \bgroup\vfil\aftergroup\estparrow%
     \let\temp=}%
  \def\estparrow{\vfil\egroup%
     \ht0=\ht\stablestrutbox%
     \dp0=\dp\stablestrutbox%
     \wd0=\stablesdummy%
     \box0}%
  \def|{\global\advance\stablesmode by 1&&&}%
  \def\|{\global\advance\stablesmode by 1&\omit\vrule width 0pt%
         \hfil&&}%
  \def\vt{\global\advance\stablesmode by 1&\omit\vrule width \stablesthinline%
          \hfil&&}%
  \def\vtt{\global\advance\stablesmode by 1&\omit\vrule width
\stablesthickline%
          \hfil&&}%
  \def\vttt##1{\global\advance\stablesmode by 1&\omit\vrule width ##1%
          \hfil&&}%
  \def\vtr{\global\advance\stablesmode by 1&\omit\hfil\vrule width%
           \stablesthinline&&}%
  \def\vttr{\global\advance\stablesmode by 1&\omit\hfil\vrule width%
            \stablesthickline&&}%
  \def\vtttr##1{\global\advance\stablesmode by 1&\omit\hfil\vrule width ##1&&}%
  \stableslines=0%
  \stablesomitfalse}
\def\stablesdef{\bgroup\stablestrut\borderrule##\tabskip=0pt plus 1fil%
  &\stablesleft##\stablesright%
  &##\ifstablesright\hfill\fi\internalrule\ifstablesright\else\hfill\fi%
  \tabskip 0pt&&##\hfil\tabskip=0pt plus 1fil%
  &\stablesleft##\stablesright%
  &##\ifstablesright\hfill\fi\internalrule\ifstablesright\else\hfill\fi%
  \tabskip=0pt\cr%
  \ifstablesborderthin%
    \thinline%
  \else%
    \thickline%
  \fi&%
}%
\def\endtable{\advance\stableslines by 1\advance\stablesmode by 1%
   \message{- Rows: \number\stableslines, Columns:  \number\stablesmode>}%
   \stablesel%
   \ifstablesborderthin%
     \thinline%
   \else%
     \thickline%
   \fi%
   \egroup\stablesend%
\endgroup%
\global\stablesinfalse}
%
%
\chapter{Introduction}

In [\moore], bosonic and heterotic strings were considered in toroidal backgrounds that included a compact time coordinate, 
and the corresponding T-duality symmetries were investigated. Although such backgrounds can have pathological features
associated with closed timelike loops, 
they can be useful in studying the symmetries of string theory. The more dimensions that are compactified, the
larger the duality symmetry [\CJ-\HT], and    a background in which all space and time dimensions are compactified would 
give a phase with a huge duality symmetry [\julia-\HT].
Moreover, there are clearly solutions of string theory which have compactified time, and studying these can give important
information about string theory.
Branes wrapped both around compact time and compact space dimensions have recently been considered in [\gibrap].

Compactification of the bosonic string on a spatial $n$ torus $T^n$ gives a theory with a moduli space
$$ {O(n,n)\over O(n)\times O(n)} \times \R^+
\eqn\abc$$
and a T-duality group $O(n,n;\Z)$ while compactification on a Lorentzian torus $T^{n-1,1}$ with $n-1$ spacelike  circles and
one timelike one gives a moduli space
$$ {O(n,n)\over O(n-1,1)\times O(n-1,1)} \times \R^+
\eqn\abc$$
and the T-duality  group remains $O(n,n;\Z)$.
T-duality in a timelike direction swaps the time component $P^0$ of the momentum with the number of times a string wraps
around the timelike circle and so has some unusual features. Nevertheless, the arguments of [\buscher-\TD] show that
it should be a perturbative symmetry for bosonic or heterotic strings in  a background with a timelike Killing vector
with compact orbits, as well as for spacelike ones.

In [\HJ], and in [\CPS,\Stelle], the compactification of type II theories on Lorentzian tori
$T^{n-1,1}$
 was considered, and it was found that the U-duality group remained the same as that for compactification on a
spacelike torus
$T^n$, but the moduli space, which was a  coset space $G/H$ for the $T^n$ reduction, became $G/\ti H$ where $\ti H$
is a certain non-compact form of
$H$. One might expect that timelike T-duality in type II theories should give a straightforward generalisation of
that for the bosonic and heterotic strings but, as we shall see, this is not the case and there are a number of
surprises. This is the first of a series of papers investigating timelike T-duality and its consequences in type II
string theory and in M-theory; one of the motivations is to further investigate the various limits of M-theory and
the web of dualities linking them. 

It will be argued that the timelike T-duals of the IIA and IIB string theories are new theories that will be referred to as the
$IIB^*$ and $IIA^*$ string theories, respectively. It should be stressed that these theories can be written down directly in
9+1 non-compact dimensions; timelike compactifications are only used to link these with the 
type II theories, so that the IIA (IIB) theory on a timelike circle of radius $R$ is equivalent to the
$IIB^*$ ($IIA^*$) theory on a timelike circle of radius $1/R$. If, for some reason,  the type II string theories could not be
formulated in a toroidal background with a timelike circle, then the type II and the type $II^*$ theories would be distinct.
As will be seen, the type $II^*$ theories have a number of problems, and only some of these   will be addressed in this paper.
However, if timelike compactifications of type II strings are consistent, then the type $II^*$ theories should  be
consistent also, at least when compactified on a timelike circle.

For compactification of type II theories on a spacelike circle, T-duality identifies type IIA on a spacelike circle of
radius
$R$ with type IIB on a circle of radius $1/R$ [\ddua,\dsei],  but the IIB theory is {\it not } the timelike T-dual
of the IIA theory [\CPS].
 Consider
type IIA compactified on a timelike circle of radius 
$R$,  which gives a theory in 9 Euclidean dimensions.
The limit in
which $R \to 0$  should give a decompactification to a dual string theory in 9+1 dimensions, as in the bosonic case,
and the IIA theory with radius $R$ should be  equivalent to the 
dual theory  on a timelike circle of radius  $1/R$. The limit $R \to 0$ corresponds to taking M-theory on a
Lorentzian torus $T^{1,1}$ in the limit in which both radii shrink to zero size.
The moduli space of $T^{1,1}$ is 
$${ SL(2,\R)\over SO(1,1) }\times \R ^+
\eqn\abc$$ 
and the limit should give a theory in 9+1 dimensions with scalars taking values in 
$${ SL(2,\R)\over SO(1,1) } 
\eqn\tobsr$$ 
This is different from the usual IIB coset space 
$${ SL(2,\R)\over SO(2) } 
\eqn\abc$$ 
and so the theory cannot be the usual IIB theory. We will refer to the timelike T-dual of the IIA theory as the type $IIB^*$
theory and to the timelike T-dual of the IIB theory as the type $IIA^*$
theory. 

The $IIB^*$ theory has  scalar coset space \tobsr, and the RR scalar has a  kinetic term of the
wrong sign. In fact,  the bosonic sectors of the $IIA^*$ and $IIB^*$ supergravities are, as will be
seen in section 4, obtained from those of the IIA and IIB theories by continuing all the RR $n$-form
gauge fields $C_n \to -i C_n$, so that the kinetic terms of all the RR gauge fields have the wrong
sign. The full type $IIA^*$ and $IIB^*$ string theories have a twisted form of $N=2$ supersymmetry
and are
 obtained by acting on the type IIA and type IIB  string theories with $i^{F_L}$, where $F_L$ is the
left-handed fermion number. The NS-NS sector of the supergravity is unchanged by this duality, but
the RR gauge fields all become ghost-like, and the resulting supergravity theories have ghosts.  In
[\gibras], similar actions were considered in which kinetic terms of the wrong sign signalled
instabilities. However,    when   type  $II^*$ string theory is compactified on a timelike circle,
it is equivalent to a conventional type II theory on the dual timelike circle, and so it is no more
pathological than   the  timelike compactification  of this conventional theory.

Ghosts occur generically for timelike compactification of gauge theories. For example, Yang-Mills
theory compactified on time gives a Euclidean theory with a scalar from the reduction of $A_0$ with
a  kinetic term of the wrong sign, but in the full higher dimensional theory, $A_0$  can be gauged
away, at least in the topologically trivial sector without Wilson lines in the timelike
direction. Thus if all
the Kaluza-Klein modes are kept, the theory should be ghost-free (in the topoliogically trivial
sector) as a result of the higher-dimensional gauge invariance, but   if the Kaluza-Klein modes are
truncated the resulting theory has a scalar ghost. It is sometimes possible to twist such a
Euclidean-space gauge theory obtained via a timelike reduction to obtain a well-defined topological
theory [\Bob,\thom].  These matters will be discussed further later.  However, the situation for the
type
$II^*$ theories might be similar. If the type $II^*$ string theories are truncated down to their 
supergravity limits, the supergravity theories have ghosts. However, in the full string theories, it
is possible that the string gauge symmetries can be used to eliminate  the ghosts.  Indeed,  the
type $II^*$ theories are linked by T-duality to the type II theories which are ghost-free, at least
perturbatively.

It is possible that, for some reason,   closed timelike curves are not allowed in string theory or M-theory,
and there is a stringy
\lq chronology protection principle' analogous to that proposed in [\hawtim], in which case the theories obtained by
postulating a compact time and performing a timelike T-duality might make the inconsistencies or instabilities more
apparent. Even in that case, there would be topological twisted versions of these theories that should be consistent, as we
will see. However,  it is clear that if timelike compactifications of type II theories are
contemplated, one is led ineluctably to consider type $II^*$ theories.
 
The type $II^*$ string theories exist in their own right in $9+1$ dimensional Minkowski space, irrespective of whether or not  closed timelike curves  are  \lq allowed'.
However, the link to the type II theories through timelike T-duality guarantees that much of the formal structure of the type II theories immediately carries over to the type $II^*$ theories, which might be thought of as a different real form of an underlying 
complexified theory. In particular, the  type $II^*$ theories will be supersymmetric and many of their properties
can be found by simply tracing the sign changes that come about in going from the type II to the type $II^*$ theories.

Type II theories have D$p$-branes corresponding to imposing Dirichlet boundary conditions in $9-p$ spatial dimensions
$x^{p+1},..., x^{9-p}$ and Neumann conditions in the remaining dimensions $t,x^1,..., x^p$, so that the ends of the string
are confined to a $p+1$ dimensional timelike D$p$-brane parameterised by  $t,x^1,..., x^p$ [\ddua]. A T-duality in
the time direction changes the boundary conditions in $t$   from Neumann to Dirichlet, so that the end of the
string is confined to a  spacelike surface at fixed $t,x^{p+1},..., x^{9-p}$
parameterised by $x^1,..., x^p$. We will refer to this $p$-dimensional spacelike surface as an E$p$-brane; for example, an
E$0$-brane is at a fixed point or event in space and time and so is a Minkowski space analogue of a D-instanton
[\dinst,\sevbrane]. An E$1$-brane can be thought of as the spacelike world-line of a tachyon, 
so that whereas a D$0$-brane is associated with a particle, an E$1$-brane is associated with a tachyon. 
The E$n$-branes with $n>1$ can be thought of as higher dimensional analogues of a tachyon and the corresponding supergravity solutions might be thought of as the field configuration 
due to such tachyons. As tachyons are associated with vacuum instability, it might be expected that the E-branes might signal some instability; this will be addressed in section 11.
Note, however, that gauged supergravities have tachyons (particles with negative mass squared) but that nonetheless anti-de Sitter space solutions can be completely stable and the space-time curvature tames the potential instabilites [\adsstab].

 The
type
$II^*$ theories do not have D-branes, but have E-branes instead. We will be careful to distinguish
between E$p$-branes, which occur in the Lorentzian signature type $II^*$ theories, and  D-instantons (and their
D$(p+1)$-instanton generalisations [\dinst]), which occur in the Wick rotated theory.

The world-volume theory of an E$p$-brane is the $p$-dimensional  Euclidean \sym\ theory obtained by compactifying the usual
9+1 dimensional \sym\ theory on $T^{9-p,1}$. In particular, $N$ coincident  E$4$-branes  gives a superconformal theory in 4
Euclidean dimensions with $U(N)$ gauge symmetry, and arguments similar to those of Maldacena [\mal] lead, as will be shown in
sections 8,9 and 10, to the conjecture that the large $N$ limit should be equivalent to type $IIB^*$ string theory in a 5
dimensional de Sitter space background (so that the 4-dimensional Euclidean conformal group $SO(5,1)$ becomes the
5-dimensional de Sitter group). Moreover, the Euclidean \sym\ theory can be twisted to obtain a topological field theory
[\Bob,\thom], so that there should be a corresponding twist of the type $IIB^*$ string theory to obtain a topological string
theory, which is dual to the large $N$ limit of the $U(N)$ topological \sym\ theory. The physical interpretation of  
E-branes will be considered in section 11.

\chapter{Euclidean Superconformal Symmetry and Super Yang-Mills Theory}

The field content of super Yang-Mills in 9+1 dimensions is a vector field $A_M$ and a 
Majorana-Weyl spinor $\ll $, both of which are Lie-algebra-valued. Dimensional reduction on a
spacelike torus
$T^n$ gives a theory in $10-n$ dimensions with a vector $A_\mm$ and $n$ scalars $\ff ^i$ from the
internal components $A_i$ of $A_M$. There is an $SO(n)$ R-symmetry arising from the
original 10-dimensional Lorentz symmetry under which the scalars transform as a vector and the
fermion fields transform according to a spinor representation. 

Euclidean Yang-Mills theory in ten dimensions
 can similarly be reduced on $T^n$ to give a vector and $n$ scalars with an $SO(n)$ symmetry
and this can be viewed as the Wick rotation of the bosonic sector  of the corresponding 
super Yang-Mills theory. However, the Wick rotation of the fermion sectors is problematic; for
example, there is no Majorana-Weyl spinor in ten Euclidean dimensions.
If 
  a Weyl or Majorana spinor is used, then the number of fermion fields is doubled
and there is no supersymmetry. 
One approach to Euclidean quantization is that of
  Osterwalder and Schrader
 [\OS].   Dimensionally reducing the fermion-doubled theory again gives a fermion-doubled
theory in $10-n$ dimensions with the fermions transforming as a spinor of $SO(n)$. There are a
number of ways of dealing with this in the quantum theory and the Euclidean functional integral --
see [\vanwick] for a recent discussion  and list of references --
 but 
there is usually no supersymmetric Euclidean action; for example, the
Wick rotation of the bosonic sector of the \sym\ theory has no supersymmetric completion.

Alternatively, one can reduce the 
9+1 dimensional \sym\ theory on one time and $n-1$ spatial dimensions 
to obtain a  theory in $10-n$ Euclidean dimensions. 
This has 16  supersymmetries and an $SO(n-1,1)$ R-symmetry under which the scalars transform as a
vector and the fermions as  a spinor.
The scalar from the time component $A_0$  of the vector potential has a kinetic term of the wrong
sign, so that the theory has ghosts. 

Thus in $D=10-n$ Lorentzian dimensions there is the usual  \sym\ theory with 16 supersymmetries and
$SO(n)$ R-symmetry (or more properly, the double cover $Spin(n)$), and any Wick rotation of this
would be a theory in
$D$ Euclidean dimensions, again with
$SO(n)$ R-symmetry. However, the fermion sector and supersymmetry are problematic. There is also a Euclidean theory in $D$ dimensions obtained by timelike reduction, with $SO(n-1,1)$
R-symmetry and 16 supersymmetries. We will refer to the Wick-rotated theory as \lq Euclideanised'
and the timelike reduction as \lq Euclidean'.

For example, in $D=4$, the Lorentzian $N=4$ theory has $SU(4)\sim SO(6)$ R-symmetry,
the Euclideanised theory again has an $SU(4)$ internal symmetry, while the Euclidean theory
has $SO(5,1)$ R-symmetry and is invariant under an $N=4$ Euclidean
super-Poincar\'e algebra under which the 4 supercharges transform as a ${\bf 4 }$ 
of $SO(5,1)$.
The spinorial generators
arise as  $SU(2)$-pseudo-Majorana-Weyl
spinors of $SO(5,1)$, in the terminology of [\KT].

The Euclidean theory arising from a timelike reduction and a truncation of the Kaluza-Klein modes
gives a non-unitary theory, but if the full Kaluza-Klein spectrum is kept, the theory is the
original unitary gauge theory, albeit on a background that includes a timelike torus, $T^{n-1,1}$.
In particular, one can choose a physical gauge in which the longitudinal component $A_0$ is
eliminated locally, so that the scalar with a kinetic term of the wrong 
sign is removed (locally). The full theory  is ghost-free (at least in the sector without timelike
Wilson lines) and it is the truncation of the Kaluza-Klein modes that leads to a theory with
ghosts.

There are three possible  twistings  of $N=4$ \sym\  to give topological field
theories [\yam,\vafwit]. The nature of these twistings has recently been clarified in [\Bob,\thom],
where it was shown that these theories are best understood as twistings of the $D=4$  Euclidean
\sym\ theory
with $SO(5,1)$ R-symmetry and $SO(4) \sim SU(2)_L \times SU(2)_R$ Lorentz symmetry   [\Bob,\thom].
The  three   distinct topological theories 
 arise from twisting the 
the Lorentz symmetry with a subgroup of the R-symmetry.
Twisting the 
$SO(4)$ Lorentz symmetry with
$SO(4)
\subset SO(5,1)$ gives the B-model with an 
$SO(1,1) $ internal symmetry, twisting the $SU(2)_L $ subgroup of the Lorentz symmetry with the
subgroup 
$SU(2)_A$ of the R-symmetry, with the embedding $SU(2)_A\times SU(2)_B\times SO(1,1) \subset
SO(5,1)$, gives the half twisted model with
$SU(2)\times SO(1,1)$ internal symmetry, while twisting the $SU(2)_L $ subgroup of the Lorentz
symmetry with the subgroup 
$SO(3)$ of the R-symmetry, with the embedding $SO(3)\times SO(2,1) \subset SO(5,1)$, gives the
A-model with $SO(2,1)$ internal symmetry. The actions for all of these topological theories contain
fields with kinetic terms of the wrong sign.

For the \sym\ theory in  $D$ dimensions on a $D$-manifold $M$, there is a quasi-topological theory if $M$ has special
holonomy so that it 
admits   commuting Killing spinors $ \aa$,
which can be used to construct a  nilpotent   BRST operator,
so that  
the physical states are defined to be BRST cohomology classes.
These give higher dimensional analogues of topological field theories  [\Bob,\thom,\sing].

The usual \sym\ theory in 2+1 Lorentzian dimensions has an $SO(7)$ R-symmetry and dualising the vector to
an extra scalar gives a theory with 8 scalars and an $SO(8)$ R-symmetry, which is the M$2$-brane world-volume theory. This has a conformal fixed
point at which the Poincar\' e symmetry is enlarged to the 3-dimensional conformal group $SO(3,2)$ [\Seib].
The full supersymmetric theory is superconformally invariant, invariant under the superconformal
group
$OSp(4/8)$, whose bosonic subgroup is $SO(3,2)\times SO(8)$.
The Euclideanised version of this should   have $SO(4,1)\times SO(8)$
symmetry (with $SO(4,1)$ the conformal group in 3 Euclidean dimensions)
but, as was to be expected, there is no superconformal algebra containing this.
The Euclidean \sym\ theory obtained by reduction on $T^{6,1}$, however, has 7 scalars, one vector and an $SO(6,1)$ R-symmetry.
Dualising the vector gives
an extra scalar  (with kinetic term of the wrong sign). This results in a theory  with 16
supersymmetries and
$SO(6,2)=SO^*(8)$ R-symmetry. 
There are 8 scalars transforming as a vector of $SO(6,2)$, and two of the scalars are ghosts. 
This theory should also have a  conformal point  at which it is   invariant under $SO(4,1)\times SO(6,2)$.
lt is also supersymmetric and so
is in fact   superconformally invariant, with  
a superconformal group whose bosonic subgroup is  $SO(4,1)\times SO(6,2)$ and which was found in [\vans], where it
was named
$OSp^*(4/8)$, as it is a different real form of  $OSp(4/8)$. (There is no superconformal group whose
bosonic subgroup is  $SO(4,1)\times SO(7,1)$  [\vans], confirming the sign of the extra scalar's kinetic
term.)

The $N=4$ \sym\ theory in  3+1 Lorentzian dimensions  is
superconformally invariant, and the superconformal group is $SU(2,2/4)$, whose bosonic subgroup is
the product of the   $SO(6)\sim SU(4)$ R-symmetry and the $SO(4,2)\sim SU(2,2)$ conformal group in
3+1 dimensions.
The Euclideanised version (from Wick rotation) has $SO(5,1)\times SO(6)$ symmetry
(with $SO(5,1)$ the conformal group in 4 Euclidean dimensions),    and played an
important role in [\witt]; however, there is no Euclidean superconformal group in this case.
 The Euclidean
version (from timelike dimensional reduction), however, 
is conformally invariant and supersymmetric and so must be invariant under a superconformal symmetry whose bosonic
subgroup is $SO(5,1)\times SO(5,1)$ and which contains the super-Poincar\' e algebra. 
It is straightforward to derive it as the symmetry algebra of the Euclidean \sym, and is 
  a different real form of $SU(2,2/4)$, which will be
denoted
$SU^*(4/4)$; it  was first given in [\luk].

\chapter{Euclidean D-branes and \sym}

In type II string theory one can have open strings that satisfy Neumann boundary conditions 
in $p+1$ dimensions $X^\mm$, $\mm =0,1,..., p$, and Dirichlet boundary conditions in the remaining
$9-p$ dimensions $X^i$, $i=1,...,9-p$. (Here $0\le p\le 9 $ and $p$ is odd for type IIB and even for
type IIA theories). These constrain the ends of the strings to lie in a $p+1$ surface $X^i=constant$,
which is a D$p$-brane [\ddua].  These are dynamical extended \lq solitons'. T-duality 
in a transverse  direction $X^{i_1}$ (assumed to be compact) changes the $X^{i_1}$ boundary condition
from  Dirichlet to Neumann, taking the D$p$-brane to a D$p+1$ brane, and conversely
a T-duality in a spatial longitudinal direction takes 
a D$p$-brane to a D$p-1$ brane

In the Wick-rotated or Euclideanised string theory, Dirichlet boundary conditions in $10-q$
(with $0\le q \le 10$)
dimensions gives a $q$-dimensional Euclidean surface which is interpreted as an extended instanton
[\dinst]. These Euclideanised D-branes are sometimes referred to as $q$-instantons or  (confusingly) as D$q-1$
branes, so that an  instanton which is pointlike in 10 Euclidean dimensions is a $p$-brane with $p=-1$. These instantonic
branes give important contributions to the Euclidean functional integral [\beck,\dinstcal].

However, it is also consistent (formally) to consider boundary conditions for  
string theory with Lorentzian signature in which Dirichlet conditions are imposed in the time direction.
Then open strings   satisfy Neumann boundary conditions 
in $p $ dimensions $X^\mm$, $\mm =1,..., p$, and Dirichlet boundary conditions in the remaining
$10-p$ dimensions $X^0,X^i$, $i=1,...,9-p$, giving a spacelike $p$-dimensional surface 
$X^i=constant$, $X^0=constant$ at a fixed instant in time. 
  These Euclidean
$p$-branes will be referred to here as E-branes, so as to distinguish them from D-branes. 
An E$0$-brane, for example, is located at a
point in 10-dimensional space-time.
Although the interpretation of such branes is unclear, they occur in the theory and one of the aims
of this paper will be to study some of their properties.

If the time dimension is compact,   a T-duality in the time direction  
  takes a D$p$-brane to an E$p$-brane, and an E$p$-brane to a D$p$-brane.
Thus if T-duality   in a compact time-direction 
is allowed, as is usually
supposed, then there must be E$p$-branes in the theory. 
 T-duality in a longitudinal direction takes an E$p$-brane to an E$(p-1)$-brane while
T-duality in a spacelike transverse direction takes an E$p$-brane to an E$(p+1)$-brane.
E$p$-branes with $p$ even occur in the $IIB^*$ theory obtained from type IIA via a T-duality  on a timelike  circle, while
E$p$-branes with $p$ odd occur in the $IIA^*$ theory obtained from type IIB from a timelike  T-duality.

The dynamics of a D-brane is given in terms of the strings ending on it, so that the low-energy
effective dynamics are given by the zero-slope limit of these strings, which is $p+1$
dimensional \sym\ on the brane. In particular, the scalar fields are collective
coordinates for the brane position. Similarly, the effective description of D-branes in Wick-rotated
string theory is   Euclideanised or  Wick-rotated
\sym, while that of  an E$p$-brane  is the Euclidean \sym\ theory obtained by reducing 
from $9+1$ dimensions in $9-p$ spatial and one time dimensions. The ghost scalar field is the
collective coordinate for the time direction.

\chapter{Timelike T-Duality and Type II  Theories}

IIA string theory compactified on a spacelike circle of radius $R$ is T-dual to IIB string theory
compactified on a circle of radius $1/R$ [\ddua,\dsei]; the two type II theories are different
decompactification limits of the same theory. The low-energy limits of the two string theories are
the type IIA and type IIB supergravity theories and the T-duality is reflected in the fact  that
the dimensional reduction of the two supergravities to 9 dimensions   give the same
9-dimensional supergravity theory [\bergort].
However, whereas for  supergravity compactified on a  circle of radius $R$, the limit $R \to 0$ gives
a truncation to nine-dimensional supergravity, the corresponding limit for string theory is a
decompactification to the T-dual string theory.

Consider now the case of timelike reductions [\HJ,\CPS] and T-duality.
For the bosonic and heterotic strings, timelike T-duality is straightforward (at least
perturbatively), and takes the theory on a circle to the same theory on a dual circle [\moore]. This is
straightforward to check by manipulations of the functional integral, and the transformations of the
background metric, dilaton and 2-form fields are those of [\buscher,\rocver] (with the shift
of the dilaton   proportional to $\log \vert k^2 \vert$, where $k^2$ is the length squared of the Killing vector in
the direction being dualised).  This T-duality is reflected in the
$U(1)$ symmetry of the dimensionally reduced 9-dimensional  classical field theory under which the
Maxwell fields from the metric and the 2-form field are rotated into one another. In the quantum
theory, this is broken to a  discrete $\Z _2 $ subgroup, the T-duality.

For the type II theory, the situation is more complicated. 
The timelike reduction of the type IIA and type IIB supergravities gives two {\it different}
9-dimensional Euclidean supergravities [\CPS]. 
Indeed,   type IIB on a timelike $S^1$ has scalars taking values in 
$SL(2,\R)/ SO(2) \times \R ^+$, where the  modulus in $\R ^+$ is  the radius of the compactifying circle,
 while IIA on a timelike $S^1$ is the same as 
11-dimensional supergravity compactified on a Lorentzian torus $T^{1,1}$
and so will have scalars taking values in $SL(2,\R) / SO(1,1) \times \R ^+$, the moduli space of $T^{1,1}$.

However, the type IIA theory on a timelike circle of
radius
$R$  should be T-dual to some string theory on  a timelike circle of
radius
$1/R$, so that the limit $R \to 0$ should give a T-dual string theory in 9+1 non-compact dimensions.
The timelike T-duality cannot then take the IIA on a timelike circle to IIB on the dual timelike circle 
[\CPS], as we will
see below.   Here we will   assume that timelike T-duality does make sense in type II theories
and  see  where this leads.

The bosonic action of the IIA supergravity is
$$\eqalign{
S_{IIA}= 
\int
 d^{10} x &
\sqrt{-g}\left[
 e^{-2 \Phi} \left( R+ 4(\partial   \Phi )^2  
-H^2 
\right)\right.
\cr &
\left. -G_2^2 - G_4^2 \right] + {4\over \sqrt 3}\int G_4 \wedge G_4 \wedge B_2 + \dots
\cr}
\eqn\twoa$$
while that of IIB supergravity is
$$
\eqalign{S_{IIB}= 
\int
 d^{10} x &
\sqrt{-g}\left[
 e^{-2 \Phi} \left( R+ 4(\partial   \Phi )^2  
- H^2 
\right)\right.
\cr &
\left.-G_1 ^2 -  G_3 ^2 - G_5 ^2 \right] + \dots
\cr}
\eqn\twob$$
Here  $\Phi $ is the dilaton, $H=dB_2$ is the field strength of the NS-NS 2-form gauge field $B_2$ and 
$G_{n+1}=dC_n+\dots$ is the field strength for the RR $n$-form gauge field $C_n$. The field
equations derived from the IIB action \twob\ are supplemented with the self-duality constraint
$G_5=*G_5$. The dimensional reduction of either on a spacelike circle gives the type II
supergravity in 8+1 dimensions, which we will refer to as the $II_{8+1}$ theory, with bosonic action  [\bergort]
$$\eqalign{S_{II_{8+1}}= 
\int
 d^{9} x &
\sqrt{-g}\left[
 e^{-2 \Phi} \left( R+ 4(\partial   \Phi )^2  
-H^2  - d \chi ^2 - F_g^2 -F_B^2
\right)\right.
\cr & \left.-G_1^2 
-G_2^2 -G_3^2- G_4^2 
\right]+\dots
\cr}
\eqn\nin$$
where $ F_g,F_B$ are the 2-form field strengths for the NS vectors resulting from the reduction of
the metric $g_{\mu \nu}$ and the NS-NS 2-form $B_2$ respectively, and $\chi$ is the scalar arising from
the reduction of the metric.  The three scalars $\Phi,\chi,C_0$ take values in the coset
$$
{SL(2,\R)\over SO(2)}\times \R ^+
\eqn\slco$$
and the action is invariant under an $SL(2,\R)\times \R^+$ duality symmetry, broken to $SL(2,\Z)$
in the quantum theory [\HT].

Consider instead the dimensional reduction on a timelike circle, to give a theory in 9 Euclidean
dimensions.
The reduction is similar to the usual case, but the signs of some of the kinetic terms are
reversed.
 Timelike reduction of the metric gives a graviphoton with a kinetic term with a reversed
sign, together with the usual metric and scalar, while the reduction of an $n$-form gauge field
gives an $n$-form with an unchanged sign and an $n-1$ form gauge field with 
a  kinetic term with a reversed
sign.
 Reducing the type IIB theory gives a  theory with bosonic kinetic terms
$$\eqalign{S_{IIB_9}= 
\int
 d^{9} x &
\sqrt{g}\left[
 e^{-2 \Phi} \left( R+ 4(\partial   \Phi )^2  
-H^2  - d \chi ^2 + F_g^2 +F_B^2
\right)\right.
\cr & 
\left.-G_1^2 
+G_2^2 -G_3^2+ G_4^2 
\right]+\dots
\cr}
\eqn\ninb$$
and scalars taking values in \slco, while the timelike reduction of the IIA theory gives
$$\eqalign{S_{IIA_9}= 
\int
 d^{9} x &
\sqrt{g}\left[
 e^{-2 \Phi} \left(  R+ 4(\partial   \Phi )^2  
-H^2  - d \chi ^2 + F_g^2 -F_B^2
\right)\right.
\cr & \left.
+G_1^2 
-G_2^2 +G_3^2- G_4^2
\right]+\dots
\cr}
\eqn\nina$$
with scalars taking values in the   coset space
$$
{SL(2,\R)\over SO(1,1)}\times \R^+
\eqn\slcon$$
The RR scalar $C_0$ has a kinetic term of the wrong sign, resulting in a 
coset space ${SL(2,\R)/SO(1,1)}\times \R^+$ with a Lorentzian metric.
We shall refer to these as the  type $IIB_{9}$ and type $IIA_{9}$ theories, respectively;
they differ from each other in the signs of some of the kinetic terms.

The type $IIA_{9}$ theory cannot be obtained by reduction of the IIB theory, and the 
type $IIB_{9}$ theory cannot be obtained by reduction of the IIA theory, so that the IIA and IIB
theories cannot be related by a T-duality on a timelike circle.
However, the type II theories should have timelike T-duals and these should have effective
supergravity actions that dimensionally reduce to the $IIA_{9}$ and $IIB_{9}$  actions.
In the NS-NS sector, the timelike T-duality acts straightforwardly through Buscher-type
transformations, but there must be some sign changes in the RR sector.
Consider the  type $IIA^*$ and type $IIB^*$ actions given by reversing the signs of the RR kinetic
terms in \twoa,\twob\ to give
$$\eqalign{S_{IIA^*}= 
\int
 d^{10} x &
\sqrt{-g}\left[
 e^{-2 \Phi} \left(  R+ 4(\partial   \Phi )^2  
-H^2 
\right)\right.
\cr &
\left.
+G_2^2 +G_4^2 
\right] + \dots
\cr}
\eqn\twoas$$
and
$$
\eqalign{S_{IIB^*}= 
\int
 d^{10} x &
\sqrt{-g}\left[
 e^{-2 \Phi} \left( R+ 4(\partial   \Phi )^2  
- H^2 
\right)\right.
\cr &
\left. +G_1 ^2+  G_3 ^2+ G_5 ^2 \right] + \dots
\cr}
\eqn\twobs$$
where the field equations from \twobs\ supplemented by the constraint $G_5=*G_5$.
The timelike reduction of the  type $IIA^*$ action \twoas\ gives the $IIB_9$ action \ninb\ and that of
the  type $IIB^*$ action \twobs\ gives the $IIA_9$ action \nina. 
This suggests that timelike T-duality relates IIA to a $IIB^*$ theory with effective
bosonic supergravity action \twobs, and the IIB to a $IIA^*$ theory with effective
bosonic supergravity action \twoas. It is not hard to see that there are such $IIA^*$ and $IIB^*$
supergravity theories. The two supergravity theories result from coupling $N=1$ supergravity to
the two types of $N=1$ gravitino multiplets, but choosing the wrong sign for the kinetic terms of
the gravitino multiplets, and   the resulting theory can be obtained from the usual theory
by multiplying the fields in
the gravitino multiplet by $i$. This can be extended to the full string theories, with 
the $IIA^*$ ($IIB^*$) theory obtained by acting on the IIA (IIB) theory with $i^{F_L}$ where $F_L$
is the left-handed fermion number operator.

The scalars $\Phi,C_0$ of the $IIB^*$ theory take values in the coset space
$$
{SL(2,\R)\over SO(1,1)}\eqn\abc$$
and the global $N=2$ superalgebra for either the $IIA^*$ or $IIB^*$ theory is the twisted one
$$ \{Q_i,Q_j \} = \gg ^\mm P_\mm \eta _{ij}
\eqn\twi$$
where $i,j=1,2$ labels the two supercharges (which have the same chirality in the $IIB^*$ theory
and opposite chirality in the $IIA^*$ theory)
 and  $ \eta _{ij}$
is the
$SO(1,1)$ invariant metric
$diag(1,-1)$.  
The anti-commutator of the second supercharge with itself has the \lq wrong' sign.
(Similar twisted superalgebras in two dimensions were considered in [\huto].)

Thus we have been led to the construction of type $IIA^*$ and 
$IIB^*$   string theories related 
by
timelike T-duality to the usual  IIB  and IIA string theories. The truncation to the corresponding 
type $IIA^*$ and 
$IIB^*$ supergravity theories gives
 theories with ghosts, but the full type $II^*$ string theories, at least when compactified on a timelike circle, are
equivalent to  type II string theories on the dual timelike circle. The uncompactified type II theories 
are ghost-free, and a physical gauge can be chosen in which longitudinal oscillations are eliminated.
If the type II theories remain ghost-free when compactified on a timelike circle, then the type $II^*$ theories
would be
 ghost-free also, at least  when compactified on a timelike circle.
The situation would then be similar to that encountered in the case of
Yang-Mills in section 2, in which   timelike dimensional reduction and truncation
led to a theory with ghosts, but if the full Kaluza-Klein spectrum was kept, then the theory was
ghost-free. 
Backgrounds with a timelike circle 
appear to be consistent string backgrounds, and it is interesting to understand the properties  of strings in such
backgrounds, and in particular the questions of stability and unitarity. 
It is intriguing that the problems appear to be arising only in the RR sector and not in the NS-NS sector.

\chapter{Dimensional Reduction}

The dimensional reduction of   the type $IIA^*$  and type $IIB^*$ supergravity theories on a spacelike circle give  the same theory in 8+1 dimensions, which we will refer to as the $II_{8+1}^*$ theory. The bosonic
kinetic terms are obtained from
those of \nin\ by reversing the signs of the RR kinetic terms to obtain
$$\eqalign{S_{II_{8+1}^*}= 
\int
 d^{9} x &
\sqrt{-g}\left[
 e^{-2 \Phi} \left( R+ 4(\partial   \Phi )^2  
-H^2  - d \chi ^2 - F_g^2 -F_B^2
\right)\right.
\cr &\left. +G_1^2 
+G_2^2 +G_3^2+ G_4^2 
\right]+\dots
\cr}
\eqn\nins$$
In fact, the type $IIA^*$  and $IIB^*$ string  theories are related by T-duality on a spacelike circle, so
that the 
type $IIB^*$  theory on a spacelike circle of radius $R$ is equivalent to the type $IIA^*$
 theory on a spacelike circle of radius $1/R$.
Indeed, consider type IIA theory on a Lorentzian torus $T^{1,1}$ with spacelike radius $R$ and
timelike radius $T$. The moduli space of such reductions can be 
represented by the square in figure 
1, and the spacelike T-duality of the  $IIA^*$  and $IIB^*$  theories follows from that between
IIA and IIB. Compactifying the IIA (IIB) string theory on $T^{1,1}$ and taking the limit in
which the torus shrinks to zero size gives the $IIA^*$ ($IIB^*$) string theory.

\let\picnaturalsize=N
\def\picsize{3.0in}
\def\picfilename{Eucfig.eps}
\ifx\nopictures Y\else{\ifx\epsfloaded Y\else\input epsf \fi
\let\epsfloaded=Y
\centerline{\ifx\picnaturalsize N\epsfxsize \picsize\fi \epsfbox{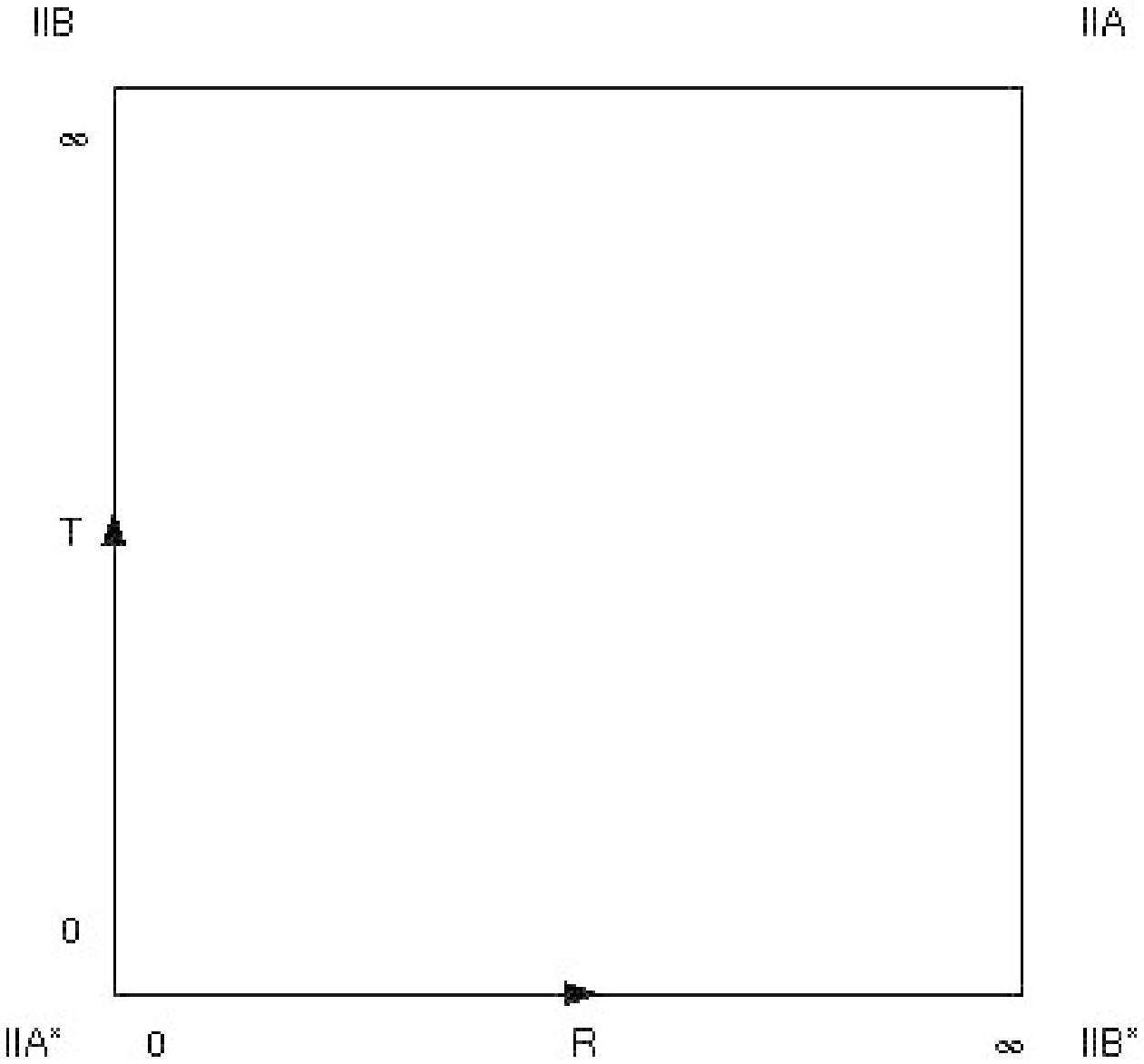}}}\fi

{\bf Figure 1}  The moduli space for the type IIA theory compactified on a Lorentzian torus $T^{1,1}$ with spacelike
radius $R$ and timelike radius $T$.

\vskip .5 cm
The standard dimensional reduction of 11-dimensional supergravity [\CJ,\julia]
on a torus $T^d$ or of type IIA supergravity on $T^{d-1}$
gives a supergravity theory in $11-d$ dimensions which is invariant under a
rigid duality symmetry
$G_d$ and a local symmetry $H_d$; the groups $G_d, H_d$ are listed in table 1.
$H_d$
is the maximal compact subgroup of $G_d$ and the theory has scalars taking
values in the coset
$G_d/H_d$.

\vskip 0.5cm

\begintable
D=11-d |
$G_d$ | $H_d$  |  {U-Duality}  \elt
 $9$ | $SL(2,\R)\times SO(1,1)$ | $SO(2) $ | $SL(2,\Z) $ \elt
 $8$ | $SL(3,\R)\times SL(2,\R)$ |  $SO(3)\times SO(2) $ | $SL(3,\Z)\times
SL(2,\Z)$
\elt
 $7$ | $SL(5,\R)$ | $SO(5)$ | $SL(5,\Z)$ \elt
 $6$ | $SO(5,5)$ | $SO(5)\times SO(5)$ |  $SO(5,5;\Z)$ \elt
 $5$ | $E_{6(6)}$ | $USp(8)$ | $E_{6(6)}(\Z)$ \elt
 $4$ | $E_{7(7)}$ | $SU(8)$ | $E_{7(7)}(\Z)$ \elt
 $3$ | $E_{8(8)}$ | $SO(16)$ | $E_{8(8)}(\Z)$ 
\endtable

{{\bf Table 1} Duality symmetries for 11 dimensional supergravity reduced, and
 M-theory compactified to $D=11-d$ dimensions on $T^d$. The classical scalar
symmetric space is $G_d/H_d$ and the U-duality group is $G_d(\Z)$.}

\vskip .5cm

The standard dimensional reduction of 11-dimensional supergravity  
on a Lorentzian torus $T^{d-1,1}$ or of type IIA supergravity on $T^{d-2,1}$
gives a supergravity theory in $11-d$ dimensions 
with $G_d$ 
rigid  symmetry
and   local  $\ti H_d$ symmetry for  the groups $G_d, \ti H_d$ are listed in table 2.

\vskip 1cm

\begintable
D=11-d |
$G_d$ | $H_d$  | $\ti H_d$ |$ K_d $ \elt
 $9$ | $SL(2,\R)\times SO(1,1)$ | $SO(2) $ | $SO(1,1)$  | $\II $ \elt
 $8$ | $SL(3,\R)\times SL(2,\R)$ |  $SO(3)\times SO(2) $ | $SO(2,1)\times
SO(2)$
| $SO(2)\times SO(2)$
\elt
 $7$ | $SL(5,\R)$ | $SO(5)$ | $SO(3,2)$ |  $SO(3)\times SO(2) $
\elt
 $6$ | $SO(5,5)$ | $SO(5)\times SO(5)$ |  $SO(5,C)$  | $SO(5)$ \elt
 $5$ | $E_{6(6)}$ | $USp(8)$ |  $USp(4,4)$ | $SO(5)\times SO(5)$ \elt
 $4$ | $E_{7(7)}$ | $SU(8)$ | $SU^*(8)$ | $USp(8)$  \elt
 $3$ | $E_{8(8)}$ | $SO(16)$ | $SO^*(16)$ |$U(8)$
\endtable

{{\bf Table 2} Toroidal reductions of M-theory   to
$D=11-d$.  Reduction on  $T^d$ gives the scalar  coset space  $G_d/H_d$
while reduction on  $T^{d-1,1}$ gives the scalar  coset space  $G_d/\ti H_d$.
$\ti H_d$ is a non-compact form of $H_d$ with maximal compact subgroup $K_d$.}

\vskip .5cm

As dimensional  reduction of IIA,IIB,$IIA^*$ or $IIB^*$ supergravities on $T^{1,1}$ give the same
8-dimensional Euclidean supergravity theory, the reduction of 
$IIA^*$ or $IIB^*$ supergravities on $T^{d-1,1}$ with $d\ge 2$  gives the 
 Euclidean supergravity theories  listed in table 2 obtained by reducing IIA or IIB on $T^{d-1,1}$.
However, the spatial reduction of the $IIA,B^*$ theories gives a new class of supergravities.
Reduction of $IIA^*$ or $IIB^*$ on $T^n$ or, equivalently, the reduction of the $II_{8+1}^*$ theory
on
$T^{n-1}$ gives a supergravity theory whose scalars take values in $G_d/H_d^*$ with $d=n+1$.
The coset $G_d/H_d^*$ is obtained by analytically continuing all the RR scalars in 
$G_d/H_d$ to change them from spacelike to timelike directions. The NS-NS scalars take values in a coset space $J_{n+1}/K^*_{n+1}$ given by
$$ {SO(n,n) \over SO(n)\times SO(n)} \times \R^+
\eqn\abc$$
for $n\le 5$, by
$$ {SO(6,6) \over SO(6)\times SO(6)} \times {SL(2,\R) \over U(1)}
\eqn\abc$$
for $n=6$, and 
$$ {SO(8,8) \over SO(8)\times SO(8)} \times \R^+
\eqn\abc$$
for $n=7$.
Then $H_d^*$ is a different real form of $H_d$ with   maximal compact subgroup of
$K_{n+1}^*$, which is $SO(n)\times SO(n)$ for $n\le 5$. This gives the symmetries in table 3:

\vskip 1cm

\begintable
D=11-d |
$G_d$ | $H_d$  | $  H_d^*$ |$ K^*_d $ \elt
 $9$ | $SL(2,\R)\times SO(1,1)$ | $SO(2) $ | $SO(1,1)$  | $\II $ \elt
 $8$ | $SL(3,\R)\times SL(2,\R)$ |  $SO(3)\times SO(2) $ | $SO(2,1)\times
SO(2)$
| $SO(2)\times SO(2)$
\elt
 $7$ | $SL(5,\R)$ | $SO(5)$ | $SO(4,1)$ |  $SO(3)\times SO(3) $
\elt
 $6$ | $SO(5,5)$ | $SO(5)\times SO(5)$ |  $SO(4,1)\times SO(4,1) $  | $SO(4)\times SO(4) $ \elt
 $5$ | $E_{6(6)}$ | $USp(8)$ |  $USp(4,4)$ | $SO(5)\times SO(5)$ \elt
 $4$ | $E_{7(7)}$ | $SU(8)$ | $SU(4,4)$ | $SO(6)\times SO(6) \times U(1) $ \elt
 $3$ | $E_{8(8)}$ | $SO(16)$ | $SO(8,8)$ |$SO(8)\times SO(8)$
\endtable

{{\bf Table 3}  Reductions on spacelike tori $T^{d-1}$ of the type $IIA^*$   theory to
$D=11-d$.  The scalar coset space is $G_d/ H_d^*$, where
$  H_d^*$ is a non-compact form of $H_d$ with maximal compact subgroup $K_d^*$.}

\vskip .5cm

These supergravities are in Lorentzian space with signature $D-1,1$ and have a twisted superalgebra
obtained from the dimensional reduction of the twisted superalgebra \twi.
The twisted superalgebra is of the form
$$
\{ Q^i , Q^j\}= \eta ^{ij} P\cdot \gg
\eqn\abc$$
where the supercharges are 
$r$-component spinors and
the index $i=1,\dots ,s= 32/r$ labels the supercharges and transforms under an $s$-dimensional representation of the
automorphism group $H^*_d$.   The metric $\eta ^{ij}$ (which may be symmetric or antisymetric, depending on the dimension) is invariant under $H^*_d$ and  has $s/2$  
eigenvalues of $+1$ and $s/2$   eigenvalues of
$-1$. For example, for $D=3$, there are 16 2-component supercharges $Q^i$, $i=1,\dots ,16$, which transform as a 
{\bf 16} of $SO(8,8)$ and $\eta ^{ij}$ is the $SO(8,8)$-invariant metric, while for $D=5$, $i=1,...,8$ and $\eta ^{ij}$ is 
the antisymmetric matrix preserved by $USp(4,4)$.
Note that $H_d^*\sim \ti H_d$ for $d=3,6$, but even for these values of $d$, the theories in tables
2 and 3 are distinct; those in table 2 are Euclidean, while those in table 3 are Lorentzian with twisted supersymmetry.

\chapter{M-Theory and F-Theory}

The $IIA_9$ theory was obtained by compactifying the IIA theory on a timelike circle, and so can also be obtained by 
compactifying M-theory on a Lorentzian $T^{1,1}$, and the scalars in the coset space \slcon\ correspond to the  $T^{1,1}$ moduli.
Just as the IIB theory can be obtained by compactifying M-theory on $T^2$ and taking the limit in which the torus shrinks to
zero size [\asp], the $IIB^*$ theory can be obtained by taking M-theory on a   $T^{1,1}$ with radii $R_1,R_2$ and taking the limit
in which both radii tend to zero, with the $IIB^*$  coupling constant determined by the limit of the ratio $R_1/R_2$.

Compactifications of type IIB theory on a space $K$ for which the complex scalar field $\tt$ is not a function on
$K$ but is the section of a bundle can be described as a compactification of F-theory on a $T^2$ bundle $B$ over $K$ 
[\fvaf].
If $K$ is an $n$-dimensional Euclidean space, then $B$ is an $n+2$-dimensional Euclidean space, and the F-theory background
is $11+1$ dimensional (i.e. eleven space and one time dimensions). On the other hand, compactifications of type $IIB^*$ theory on
a space $K$ for which the complex scalar field $\tt$ is  the section of a bundle over $K$ can be described as a
compactification of an  $F^*$-theory on a $T^{1,1}$ bundle $B$ over $K$. If $K$ is an $n$-dimensional Euclidean space, then
 $B$ is an $n+2$-dimensional  space with Lorentzian signature, and the $F^*$-theory background
is $10+2$ dimensional, with two times.
12 dimensional theories with signature $11+1$ were considered in [\ythe], while the signature $10+2$ was considered
in [\fvaf], but in the context of F-theory, not $F^*$-theory, and in [\mythe].

\chapter{E-Branes and Type $II^*$ Strings}

The E-branes of section 3 were found from D-branes of type II string theories by a timelike T-duality, and so must occur
in the type $II^*$ string theories obtained from the type II strings by a timelike T-duality. The E$p$-branes for $p$ odd
occur in the type $IIA^*$ string theory and    the E$p$-branes for $p$ even
occur in the type $IIB^*$ string theory. 
The bosonic sectors of the type $II^*$ theories can be obtained from those of the type II theories by the field redefinition
$C_n \to C_n'=-i C_n$ of the RR $n$-form potentials $C_n$. The D$p$-brane of type II carries a real RR charge corresponding to
$C_{p+1}$, while an E$p$-brane of type $II^*$ carries a real RR charge corresponding to
$C_{p }'$. An E$p$-brane of type II, corresponding to having Dirichlet boundary conditions in time as well as $9-p$ spatial
directions in   type II string theory,  would carry an {\it imaginary} RR $C_{p }$ charge, which is then interpreted as a real
$C_{p }'$ charge of the type $II^*$ theory.

The  type II supergravity solution for  a D$p$-brane, for $p<7$, has a  metric
[\hor,\huten]
$$
 ds^2=H^{-1/2}(-dt^2+dx_1^2+\dots+dx_p^2)+H^{1/2}
(dx_{p+1}^2+\dots+dx_9^2),
\eqn\dbr$$
where $H$ is a harmonic function of the transverse coordinates 
$x_{n+1},\dots,x_9$, and the dilaton and RR potential $C_{012\dots.p}$ are also given in terms of $H$.
The simplest choice is
$$ H=c+ {q\over r^{7-p}}
\eqn\hre$$
where $c$ is a constant (which can be taken to be $0$ or $1$), $q$ is the D-brane charge and $r$ is
the radial coordinate defined by
$$r^2=\sum_{i=p+1}^9 x_i^2
\eqn\abc$$
If the time coordinate $t$ is taken to be periodic, a T-duality transformation in the $t$ direction gives an E$p$-brane
solution with metric
$$
 ds^2=H^{-1/2}(dx_1^2+\dots+dx_p^2)+H^{1/2}
(-dt^2+dx_{p+1}^2+\dots+dx_9^2),
\eqn\ebr$$
This can now be generalised to allow $H$ to be a harmonic function of $t,x_{n+1},\dots,x_9$, so that it depends on  time as
well as the spatial transverse coordinates. 
In addition to the solution with $H$ given by \hre, there are solutions 
$$  H=c+ {q\over \tt ^{8-p}} 
\eqn\ehar$$
and (for odd $p$)
$$  H=c+ {q\over \ss ^{8-p}} 
\eqn\ehars$$
where $\tt, \ss $ are the proper time and distance   defined by
$$\tt^2 =  -\ss ^2=  t^2- r^2 
\eqn\abc$$
These solutions have a potential    singularity on the light-cone $t^2=r^2$  
and  arise  as    complex solutions of the original type II supergravity, 
with $C_{12\dots.p}$ imaginary, or as   real solutions of the type $II^*$ theory, with $C_{12\dots.p}$ real.
Analogous solutions were considered in [\gibras] where they were interpreted as signalling an instability.
The E-branes preserve 16 of the 32 supersymmetries of the type $II^*$ theories.
The E-brane solutions will be discussed in more detail in later sections and in  [\huku].

The E$0$-brane is closely related to the D-instanton of [\dinst,\sevbrane].
In [\sevbrane], a  \lq Euclideanised' IIB theory ten Euclidean
dimensions was proposed in which the RR scalar $C_0$ is continued to $C_0'=iC_0$, so that the action for $C_0'$ is negative
and  the scalars
$\Phi,C_0'$ take values in an $SL(2,/R)/ SO(1,1)$ coset space.
In Einstein frame, the metric of the D-instanton is just the flat 10-dimensional Euclidean metric, while
the dilaton and RR scalar are given by
$$
e^\Phi=H, \qq C_0'= H^{-1}
\eqn\rtyr$$
in terms of a harmonic function $H(x_1,\dots,x_{10})$. The spherically symmetric  instanton arises from the choice
$$ H=1+ {q\over r^8}
\eqn\teeter$$
The E$0$-brane solution of the type $IIB^*$ theory, which  has scalars $\Phi,C_0$ taking values in  an $SL(2,\R)/
SO(1,1)$ coset space but has a Lorentzian space-time signature,
 is closely related. The   metric   is   the flat 10-dimensional {\it Minkowski} metric, while
the dilaton and RR scalar are  given by
$$
e^\Phi=H, \qq C_0= H^{-1}
\eqn\rtyrs$$
but now $H$ is a harmonic function of $t,x_1,\dots,x_{9}$, i.e. it satisfies the 10-dimensional wave equation;
a simple choice is 
$$ H=1+ {q\over \tt ^8}
\eqn\ytr$$
 An E$0$-brane arises from  Dirichlet
 conditions associated with a particular point $X$ in space-time, and there are 
solutions corresponding to scalar waves emanating from the event $X$. 

\chapter{De Sitter Space and  Large $N$ Gauge Theory }

The type IIB supergravity theory has a solution [\adssol]
$$AdS_5\times S^5
= {SO(4,2)\over SO(4,1)}\times {SO(6)\over SO(5)}
\eqn\abc$$
 so that compactification on a 5-sphere gives a gauged supergravity in 5-dimensional anti-de Sitter space $AdS_5$.
The 5-dimensional anti-de Sitter space $AdS_5$
can be represented as the hyperboloid
$$-T^2-t^2 +   x_1^2+\dots + x_4^2=a^2
\eqn\adsf$$
in $\R^6$, with the metric induced from 
$$ds^2=-dT^2-dt^2 +dx_1^2+\dots +dx_4^2
\eqn\wrew$$
The RR 4-form gauge field has a field strength that is the self-dual combination of the volume form  on 
$AdS_5$ and the volume form  on $ S^5$, and its energy-density gives rise to the negative cosmological constant.
In the type $IIB^*$ theory, the RR 4-form gauge field has a kinetic term and hence an energy-momentum tensor with the \lq
wrong' sign, and so a similar ansatz gives a solution with a {\it positive } cosmological constant, and there is a de Sitter
solution $dS_5\times H^5$, where
$dS_5$ is   de Sitter space and $H_5$ is (one of the the two sheets of) the hyperboloid ${SO(5,1)/SO(5)}$, so that
$$ dS_5\times H^5
= {SO(5,1)\over SO(4,1)}\times {SO(5,1)\over SO(5)}
\eqn\csdf$$
The de Sitter space $dS_5$ can be represented as
  the hyperboloid 
$$-T^2+Y^2 +  x_1^2+\dots + x_4^2=a^2
\eqn\dsf$$
in $\R^6$, with the metric induced from the Minkowski metric
$$ds^2=-dT^2+dY^2 +dx_1^2+\dots +dx_4^2
\eqn\mink$$
while $H^5$ arises from the hyperboloid
$$-T^2+Y^2 +  x_1^2+\dots + x_4^2=-a^2
\eqn\hfi$$
again embedded in the   6-dimensional Minkowski space with metric \mink.
The hyperboloid has two sheets, and $H^5$ is taken to be one of the two.

Reduction on $H^5$ (in the sense of [\CW]) gives a gauging of the 5-dimensional Euclidean maximal 
supergravity theory from table 3, with gauge group $SO(5,1)$ arising from the isometries of $H^5$.
This has an $SO(5,1)$-invariant de Sitter vacuum which preserves all of the supersymmetries, i.e. it
has 32 Killing spinors. The vacuum is invariant under the de Sitter supergroup
$SU^*(4/4)$ of section 2, with bosonic subgroup $ SO(5,1) \times  SO(5,1)$. As in the de Sitter supergravity constructed
in [\vans], some of the fields have kinetic terms of the wrong sign.
Alternatively, one can reduce on
$dS_5$ to obtain a Euclidean supergravity on $H^5$, which is a gauging of
 the 5-dimensional maximal  supergravity theory from table 2. It has $SO(5,1)$ gauge symmetry and a maximally symmetric $H^5$
ground state, again  invariant under $SU^*(4/4)$.
Although both $H^5$ and $dS_5$ are non-compact, it seems that these solutions can be given a Kaluza-Klein interpretation if
suitable boundary conditions are imposed. If $H^5$ is regarded as the internal space, then the natural boundary conditions
are the reflective boundary conditions on $H^5$ used in e.g. [\witt], giving rise to a discrete spectrum.

Anti-de Sitter space in five dimensions is topologically $\R^4\times S^1$ with  the  $S^1$ timelike, while its covering space
is topologically $\R^5$. It has a timelike boundary, which is topologically $S^3 \times S^1$, or $S^3 \times \R$ in the
covering space. The sphere $S^3$ at infinity is of infinite radius.
The space $H^5$ was considered in detail in [\witt], where it arose as \lq Euclidean anti-de Sitter
space'. It is topologically $\R^5$ and has a boundary which is an $S^4$ at infinity. The de Sitter space $dS_5$ is
topologically $S^4
\times \R$ with the $\R$ corresponding to time. The metric can be written as
$$ ds^2 = - dt^2 + a^2 \cosh ^2(t/a) d \www ^2_4
\eqn\abc$$
where $\www ^2_n$ is the metric on a unit $n$-sphere. Thus an $S^4$ contracts from an infinite size at $t=- \infty$ to a
minimal radius $a$ at $t=0$ and then re-expands to infinite size at $t=\infty$.
There is a spacelike infinity which is topologically $S^4$ both in the future ($t=\infty$) and in the past ($t=- \infty$), and
each $S^4$ of constant $t$ is a Cauchy surface.
The solutions of   wave equations on each of these spaces are given   by specifying boundary conditions at the appropriate
hypersurface. In $AdS_5$ and $H^5$, there is a unique
 regular solution for given asymptotic boundary values, while in de Sitter space the solutions are
determined by giving  data on any Cauchy surface, e.g. initial data at $t=- \infty$.

Maldacena has  proposed [\mal] that the IIB string theory on $AdS_5\times S^5$ is dual to ${\cal N}=4$ super-Yang-Mills theory
with
$U(N)$ gauge group, at least in the limit $N \to \infty$. Both theories have  $SU(2,2/4)$ symmetry, and  the conjecture was
motivated by considering $N$ coincident D$3$-branes. The supergravity solution given by \dbr,\hre\ with $p=3$ interpolates
between the Minkowski space and the $AdS_5\times S^5$ vacua, and $AdS_5\times S^5$ is the   geometry
 which arises in
the limits used in [\mal].

We propose here a similar duality between the type $IIB^*$ string theory on $ dS_5\times H^5$ and the Euclidean
${\cal N}=4$ super-Yang-Mills theory (arising from compactifying the usual \sym\ theory in 9+1 dimensions on $T^{5,1}$) with
$U(N)$ gauge group, at least in the limit $N \to \infty$. Both theories have $SU^*(4/4)$ symmetry, which is interpreted as a
super de Sitter group in 5 dimensions  or as a superconformal group in 4 Euclidean dimensions.
Note that both theories have ghosts.

This conjecture can be motivated by considering $N$ coincident E$4$-branes; the corresponding type
$IIB^*$ supergravity solution   breaks half the supersymmetry and interpolates between the flat space
solution and
$ dS_5\times H^5$, both of which preserve all supersymmetries, just as the D3-brane interpolates
between flat space and $AdS_5\times S^5$ [\gibtow]. Moreover, the de Sitter solution $ dS_5\times
H^5$ arises in the large
$N$ limit, as in [\mal].

\chapter{Interpolating Geometries and Large $N$ Gauge Theories}

Let us begin by recalling the geometry of the D$3$-brane solution given by 
$$
 ds^2=H^{-1/2}(-dt^2+ dx_1^2+\dots+dx_3^2)+H^{1/2}
(dr^2+ r^2 d\www _4^2),
\eqn\efopo$$
with 
$$   H=1+ {a^4\over r ^{4}}
\eqn\erter$$
When $r >>a$, $H \sim 1$, so that for large $r$ the metric approaches that of flat space.
When $r<<a$, the constant term in $H$ is negligible and the metric becomes
approximately
$$
 ds^2={r^2  \over  a^2}dx_{\vert \vert} ^2  
 -{a^2 \over r^2} d r^2+a^2 d  \www _5^2
\eqn\abc$$
where 
the longitudinal metric is $dx_{\vert \vert} ^2  =-dt^2+ dx_1^2+\dots+dx_3^2$.
This metric is then that of a product of a 5-sphere of radius $a$ and 5-dimensional anti-de Sitter space with
\lq anti-de Sitter radius' $a$ and metric
$$
 ds^2={r^2  \over  a^2}dx_{\vert \vert} ^2  
 -{a^2 \over r^2} d r^2
\eqn\abc$$
Thus the D$3$ brane interpolates between  flat space and the $AdS_5\times S^5$ solution.

A similar argument shows that the E$4$-brane interpolates between flat space and the $dS_5\times H^5$ solution of the $IIB^*$
theory. The
$E4$-brane solution
\ebr,\ehar\ can be written as
$$
 ds^2=H^{-1/2}(dx_1^2+\dots+dx_4^2)+H^{1/2}
(-dt^2+dr^2+ r^2 d\www _4^2),
\eqn\efo$$
with 
$$   H=c+ {q\over \tt ^{4}}
\eqn\abc$$
where $  \tt $ is the proper  time 
$   \tt^2 =  t^2-r^2 
$. 
Taking $c=1$, the metric approaches the flat metric  far away from the light-cone $r^2=t^2$ when $\tt ^4$ is very large, while
near the light-cone the constant term in $H$ is negligible and, as we will see, the metric approaches that of the
$dS_5\times H^5$ solution. This is non-singular, and  the E$4$-brane solution is itself  non-singular, as we shall see.
The light-cone $r^2=t^2$ divides the space into two regions, and we will consider each separately.

Consider first the region $t^2 > r^2$.
Then  taking
$$ H^{1/2}= {a^2\over \tt ^2}
\eqn\htim$$
with $a^4 =q$ and
defining the Rindler-type coordinates $\tt,\bb$ by
$$
t= \tt \cosh \bb, \qq r =  \tt \sinh \bb 
\eqn\trtryru$$
 the metric \efo\ becomes
$$
 ds^2={\tt ^2  \over  a^2}dx_{\vert \vert} ^2  
 -{a^2 \over \tt^2} d \tau^2+a^2 d\ti \www _5^2,
\eqn\abc$$
where
$$ d\ti \www _5^2= d \bb ^2 + \sinh ^2 \bb  d \www _4^2
\eqn\abc$$
is the metric on $H^5$ of \lq radius'  $1$, and the longitudinal metric is now
$ dx_{\vert \vert} ^2 =dx_1^2+\dots +dx_4^2$. The metric 
$$ds^2={\tt ^2  \over  a^2}dx_{\vert \vert} ^2  
 -{a^2 \over \tt^2} d \tau^2
\eqn\abc$$
is   the de Sitter metric of  \lq radius'  $a$.

In the region $r^2> t^2$, we use   \efo\
with 
$$   H^{1/2}=  {a^2\over \ss ^{2}}
\eqn\hspa$$
($\ss ^2=r^2-t^2$) and define the  coordinates $\ss,\aa$ by
$$
r= \ss \cosh \aa, \qq t =  \ss \sinh \aa 
\eqn\dssad$$
so that the metric \efo\ becomes the $dS_5\times H^5$ metric
$$
 ds^2={\ss ^2\over a^2}  dx_{\vert \vert} ^2  
 +{a^2 \over \ss^2} d \sigma^2+a^2 d\hat  \www _5^2
\eqn\ebrsp$$
where
$$ d\hat \www _5^2= -d \aa ^2 + \cosh ^2 \aa  d \www _4^2
\eqn\abc$$
is the de Sitter metric of radius $1$, and 
the metric 
$$ds^2={\ss ^2\over a^2}  dx_{\vert \vert} ^2  
 +{a^2 \over \ss^2} d \sigma^2
\eqn\abc$$
is   the   metric on $H^5$ of radius $a$.
The light-cone $r^2=t^2$ is the boundary of $H^5$ at $\ss=0$, and is at an infinite distance from
any interior point.

In either case, the limit $c=0$ describes the  geometry near the light-cone $r^2=t^2$, while far away from the
 light-cone $ \vert t  ^2 -r^2 \vert >>a^2$,  $H \sim 1$ and the metric approaches that of flat space.  
Thus  the two regions of the
E4-brane solution both interpolate between the 10-dimensional Minkowski metric and the $dS_5\times H^5$ metric, but in one case
the interpolation is spacelike and in the other it is timelike. 

The coordinates used above only cover part of the relevant spaces. In the  case $a=0$,
we have flat 10-dimensional Minkowski space and 
 the Rindler-type 
coordinates $\tt,\bb,\dots $ or $\ss, \aa , \dots$ only cover either $\R ^4 \times C_{int}$ or $\R
^4 \times C_{ext}$, where $C_{int}$ is the interior of the light-cone $t^2>r^2$ in 6-dimensional
Minkowski-space, and $C_{ext}$ is the exterior of the light-cone $r^2>t^2$. Neither the interior 
or the exterior region of Minkowski space is
geodesically complete, as the two regions  can be linked by spacelike geodesics.
 However, if $a\ne 0$, there is a coordinate singularity (at least) on the light-cone $r^2=t^2$, and the global structure
needs to be considered more carefully.
With $c=0$, the space is $dS_5\times H^5$, and so non-singular.

Consider first the exterior of the light-cone $t^2 <r^2$, in which the
coordinates $\aa$ and the $S^4$ coordinates cover the whole of $dS_5$, and the coordinates $\ss ,  x_{\vert \vert} $ 
with $\ss >0$ cover the whole of
  $H^5$. In this representation, $H^5$ is the half-space $\ss >0$ in $\R^5$ with boundary given by 
the hyperplane $\ss=0$ together with a point at infinity.
The light-cone $r^2=t^2$  is then  the boundary $\ss=0$ of $H^5$, which is at
infinite geodesic distance from any interior point, so that now the space is geodesically complete and there are no finite
length curves passing through the light-cone to the other region.
The same remains true for the E$4$-brane with $c=1$, so as one tries to approach the light-cone $r^2=t^2$, the space becomes
the asymptotic 
$dS_5\times H^5$ geometry and the distance to the light-cone  $\ss=0$ becomes infinite.
The E$4$-brane solution given by \ebrsp\  with
$\ss $ real and positive (i.e. $r^2>t^2$) is geodesically complete and non-singular and no other region is needed.

Similarly, consider the  interior of the light-cone $r^2 <t^2$. 
In the limit $c = 0$,
 the
coordinates $\bb$ and the $S^4$
 coordinates cover the whole of $H^5$ (which is half of the 2-sheeted hyperboloid),
 but the coordinates $\tt ,  x_{\vert \vert} $ with  $\tt >0$ cover
only   half of the de Sitter space.
There is a coordinate singularity at $\tt=0$, and 
the region near the boundary $\tt=0$ is best described by representing $dS_5$
as the hyperboloid \dsf\ in the Minkowski space with metric \mink.
Then   
$\tt=T+Y$ (by an argument similar to that in the appendix of [\mal]), so 
$\tt=0$ is the intersection of the  de Sitter hyperboloid \dsf\ with the
null hyperplane $Y+T=0$, and the region $\tt>0$  is the half of the hyperboloid  in which   $T+Y>0$  (corresponding to a
\lq steady-state universe' solution). 
Then $\tt=0$ is   a coordinate singularity, and the geometry can be continued through this to give the complete
non-singular
$dS_5\times H^5$ solution.
 The interior of the light-cone splits into two regions, the past light-cone $t<r<0$ and
the future light-cone
$0<r<t$, and 
it is natural to define the proper time so that these are the regions $\tt<0$ and $\tt>0$, so that
these correspond to the two halves of the de Sitter space, $T+Y<0$ and $T+Y>0$. For the E$4$-brane solution
with $t^2>r^2$, the region near
$t^2=r^2$ or $\tt=0$ is described by a non-singular $dS_5\times H^5$ geometry, 
and $\tt \to -\tt $ is an isometry, so that one can argue as in [\dilres] that the 
space can be continued through the coordinate singularity
at $\tt =0$. Then    the region $\tt$ real or
$t^2>r^2$ of the E$4$-brane solution is also a complete non-singular solution.

There are thus two distinct complete E$4$-brane solutions, corresponding to the interior or the exterior of the light-cone,
giving rise to a timelike or a spacelike interpolation.
In [\mal],   $N$ parallel D$3$-branes separated by distances of order $\rr$ 
were considered and   the zero-slope limit
$\aa '\to 0$  was taken keeping $r=\rr / \aa '$ fixed, so that the energy of stretched strings remained finite.
This decoupled the bulk and string degrees of freedom  leaving a theory on the brane which is $U(N)$ ${\cal N}=4$ \sym\ with
Higgs expectation values, which are of order $r$, corresponding to the brane separations.  
The corresponding D$3$-brane supergravity solution is of the form
\efopo,\erter\ and
 has charge $q=a^2 \propto Ng_s/{\aa '}^2$ where $g_s$ is the
string coupling constant, which is related to the \sym\ coupling constant $g_{YM}$ by $g_s=g_{YM}^2$.
Then as $\aa '\to 0$, $q$ becomes large and the background becomes $AdS_5\times S^5$.
The IIB string theory in the $AdS_5\times S^5$ background is a good description if the curvature $R \sim 1/a^2$ is not too
large, while if $a^2$ is large, the \sym\ description is reliable. 
In the 't Hooft limit in which $N$ becomes large while $g_{YM}^2N$ is kept fixed, $g_s \sim 1/N$, so that as $N\to \infty$,
we get the free string limit $g_s\to 0$, while string loop corrections correspond to $1/N$ corrections in the \sym\ theory.

Similar arguments apply here, with the two   E$4$-brane solutions with spacelike or timelike interpolations corresponding to
whether the separation between the  E$4$-branes that is kept fixed is spacelike or timelike.
Recall that the scalars of the \sym\ theory are in a vector representation of the $SO(5,1)$ R-symmetry, where those in the {\bf 5} of
$SO(5)\subset SO(5,1)$ have kinetic terms of the right sign and correspond  to brane separations in the 5 spacelike 
 transverse dimensions, while the remaining ($U(N)$-valued) scalar  is a ghost and corresponds to timelike separations of
the E-branes.

Consider first the case of   $N$ parallel E$4$-branes of the $IIB^*$ string theory separated by distances of order $\rr$ 
in one of the 5 spacelike transverse dimensions.
We take the zero-slope limit
$\aa '\to 0$ keeping $\ss=\rr / \aa '$ fixed, so that the energy of stretched strings remains  finite.
This gives a decoupled theory on the brane consisting of the $U(N)$ ${\cal N}=4$ Euclidean \sym, with
Higgs expectation values of order $\ss$ for the scalars  
corresponding to the spacelike separations.  The corresponding supergravity background
is the E$4$-brane with spacelike interpolation, arising from the outside of the light-cone with $\ss$ real and
positive. 
We again have  $q=a^2 \propto Ng_s/{\aa '}^2$ and $g_s=g_{YM}^2$, so that for large $N$, the system can be 
described by the $IIB^*$ string theory in $dS_5\times H^5$ if $a^2$ is large and by the large $N$ Euclidean \sym\ theory when
$a^2$ is small.
In the 't Hooft limit,  string loop corrections again correspond to $1/N$ corrections in the \sym\ theory.
In the same sense that in the anti-de Sitter correspondence, the Lorentzian gauge theory can be thought of as located at the
timelike boundary of anti de Sitter space, here the Euclidean gauge theory can be thought of as located at the boundary $S^4$
of $H^5$.

For   $N$ E$4$-branes of the $IIB^*$ string theory separated by distances of order $T$ 
in the timelike transverse dimension, 
we take the zero-slope limit
$\aa '\to 0$ keeping $\tt=T/ \aa '$ fixed.
This gives a decoupled theory on the brane consisting of the $U(N)$ ${\cal N}=4$ Euclidean \sym, with
Higgs expectation values of order $\tt$ for the scalars  
corresponding to the timelike separations.  The corresponding supergravity background
is the E$4$-brane with timelike interpolation, arising from the inside of the light-cone with $\tt$ real.
Again  for large $N$, the system can be 
described by the $IIB^*$ string theory in $dS_5\times H^5$ if $a^2$ is large and by the large $N$ Euclidean \sym\ theory when
$a^2$ is small. In this case, the Euclidean gauge theory can be thought of as 
being located at the past (or future) Cauchy surface $\tt= -\infty$ ($\tt=  \infty$).

\chapter{Euclideanised Branes and de Sitter Space}

The conjectured  correspondence between large $N$ \sym\ theory and IIB string theory in $AdS_5\times S^5$ has been made more
precise in [\witt,\poly], where correlation functions of the \sym\ theory are given by the dependence of the
IIB string theory S-matrix on the asymptotic behaviour at infinity.
 In [\witt], this is formulated in terms of the Euclideanised or Wick-rotated  theory; the Euclidean version of $AdS_5$ is
the hyperboloid
$H^5$  with boundary $S^4$ and  correlation functions of the Euclideanised \sym\ theory (with $SO(6)$ R-symmetry) on $S^4$
are related to the dependence of the  S-matrix elements of the \lq Euclideanised IIB string theory' on $H^5\times S^5$ on the
boundary conditions on $S^4$.

This can be motivated as follows.
Consider the D$q$ instanton
$$
 ds^2=H^{-1/2}(  dx_1^2+\dots+dx_q^2)+H^{1/2}
(dx_{q+1}^2+\dots+dx_{10}^2),
\eqn\dins$$
where $H$ is a harmonic function of the transverse coordinates 
$x_{q+1},\dots,x_{10}$, such as
$$ H=c+ {q\over r^{8-q}}
\eqn\hiss$$
where $c$ is a constant   and $r$ is now the radial coordinate in the transverse space  defined by
$$r^2=\sum_{i=q+1}^{10} x_i^2
\eqn\abc$$
This arises from the Wick rotation   $t\to it $  
of   the D$q-1$ brane solution \dbr.

With $c=1$, the D$q$-instanton interpolates between a flat space region where $r^2$ is large and a small $r$ region where the
constant term in \hiss\ is negligible. For $q=4$, the metric in this region is given by
$$
 ds^2=H^{-1/2}(dx_1^2+\dots+dx_4^2)+H^{1/2}
( dr^2+ r^2 d\www _5^2),
\eqn\efos$$
with 
$$   H= {a^4\over r ^{4}}
\eqn\abc$$
which can be written as
$$
 ds^2={r ^2  \over  a^2}dx_{\vert \vert} ^2  
 +{a^2 \over r^2} d r^2+a^2 d  \www _6^2,
\eqn\abc$$
 and is the metric on $H^5\times S^5$. Thus the D$4$-instanton interpolates between flat space and $H^5\times S^5$ and in the
large $N$ and zero-slope limit, as in [\mal], if $a$ is large the theory is well described by the
Euclideanised string theory in 
$H^5\times S^5$, and when $a$ is small, it is well described by the large $N$ limit of the Euclideanised \sym; this leads to
the Euclideanised version of the Maldacena duality described above.

The conjectured equivalence between the Euclidean \sym\ theory and the $IIB^*$ string theory on $dS_5\times H^5$ should
similarly have a Euclideanised   formulation. 
Yang-Mills theory in $d+1$ Lorentzian dimensions is Euclideanised by Wick rotating $t\to it $ and taking $A_0 \to -i A_0$, so
that the connection 1-form remains real and the Euclidean action is positive. 
From the \sym\ theory in $9+1$ dimensional space, we can obtain a  \sym\ theory in D-dimensional Minkowski space by reducing
on a spacelike torus $T^{10-D}$ and a
\sym\ theory in D-dimensional Euclidean space by reducing
on a timelike torus $T^{9-D,1}$, whereas the Euclideanised theory in 10 dimensions can be reduced on $T^{10-D}$
to give a theory in $D$ dimensions. 
It is natural to take this as the Euclideanisation of both of the $D$-dimensional \sym\ theories; for the one in
D-dimensional Minkowski space, this is the usual analytic continuation, while for the Euclidean theory, this amounts to 
continuing the scalar $\phi$ in $D$ dimensions arising from $A_0$ by $\phi \to -i \phi$, so that its action becomes positive
and the R-symmetry changes from $SO(9-D,1)$ to $SO(10-D)$.

The Euclidean  version of the de Sitter space $dS_5$ is the 5-sphere $S^5$; after analytically continuing the
appropriate time coordinate, there is a coordinate singularity which can be removed by identifying the Euclidean time with
period given by  the inverse of the Gibbons-Hawking temperature of de Sitter space, associated with the presence of a
cosmological event horizon [\gibhaw]. Thus the Euclidean  version of $dS_5\times H^5$ is $S^5\times
H^5$, which is the same as the Euclidean  version of
$S^5\times AdS_5$; the same space has two different Lorentzian continuations, depending on which factor in the product is
continued, and the   Euclidean field theory equivalence of [\witt] can be continued in one of two ways, to give
either the de Sitter space duality or the anti-de Sitter space duality.

The Wick rotation  $t\to it$ of the E$q$-brane solution \ebr\ gives the same D$q$ instanton solution \dins,\hiss\ as  was
obtained from analytically continuing the D$q-1$ brane, and for $q=4$ this again interpolates between flat space and 
$S^5\times H^5$. 
(Note that the inside of the light-cone has become a point, so there is only one type of Euclideanised E-brane.)
Both the Euclideanised type II  and type  $II ^*$ theories are expected to admit the same D-instanton
solutions and similarly they  should admit instantons corresponding to the Wick rotation of strings and NS 5-branes; the
NS-NS sectors are the same for both the II and $II^*$ theories and so these couple to the same NS 1-branes and 5-branes.
Thus the Euclideanisations of the type II  and type  $II ^*$ theories have the same instanton-brane spectrum
and so can be taken to be the same theory in 10 Euclidean dimensions. This will be the case if the analytic continuation
of the  type $II^*$ theories is accompanied by
   an extra continuation
$C_n\to -iC_n
$ of the RR gauge fields    so that the Euclideanised action of the matter fields becomes positive and is the same as that of
the Euclideanised IIB theory; these Euclideanised actions are discussed in an appendix. This is similar to the 
redefinition $\phi \to - i \phi$ of the scalar ghost in the Euclidean \sym\ theory above, so that the Euclidean and Lorentzian
 D-dimensional \sym\ theories have the same Euclideanised version.

Then the Euclideanised version of the duality between $IIB^*$ string theory
$dS_5\times H^5$ and 4-dimensional Euclidean \sym\ 
  is the {\it same} as the   Euclideanised version of the duality between $IIB $ string theory
$AdS_5\times S^5$ and 4-dimensional Lorentzian \sym.  In both cases,   
it a duality between the Euclideanised $IIB $ theory on $H^5\times S^5$ and 4-dimensional Euclideanised \sym, and this is the
duality that was formulated in [\witt] and for which some evidence now exists.
The same dual pair of Euclideanised theories has two different Lorentzian sections.
For the space $H^5\times S^5$, one either continue the $H^5$ to $AdS_5$ or the $S^5$ to $dS_5$; for the Eulideanised \sym\
theory given by reducing from 10 Euclidean dimensions on $\R^4\times T^6$, one can either continue one of the
coordinates in $\R^4$ to get a theory in Minkowski space with $SO(6)$ R-symmetry, or  one can   continue one of the
internal $T^6$ coordinates, to obtain a theory with
$SO(5,1)$ R-symmetry in Euclidean space; for the D$4$-instanton, one can either continue one of the longitudinal coordinates
to obtain a D$3$-brane or one can continue one of the transverse directions to obtain an E$4$-brane.

Thus a precise Euclideanised  formulation of the de Sitter duality is the same as that of the anti-de Sitter duality; it is a relation
between correlation functions of the Euclideanised \sym, and the S-matrix of the Euclideanised string theory, but this can be
continued back to a Lorentzian regime in two different ways.
Continuing back to the E$4$-brane with the spacelike interpolation naturally leads to a holographical
association of the bulk theory with the Euclidean gauge theory on the boundary $\ss=0$ of $H^5$.
For the timelike interpolation, the natural association would appear  to be with the Euclidean gauge theory on the
hypersurface $\tt=0$ of the de Sitter space.

\chapter {Interpretation of E-branes and Vacuum Instability}

In this section, the geometry and interpretation of the E$4$-brane solution will be considered further.
It is is closely related to the extreme Reissner-Nordstrom  solution in 4 dimensions considered in [\gibras].
The Euclideanised Reissner-Nordstrom geometry can be continued back to Lorentzian signature in two ways, to give either  the usual 
extreme Reissner-Nordstrom metric which interpolates between $AdS_2\times S^2$ and flat space, or to give an E-brane type 
 solution 
which interpolates between $ dS_2\times H^2$ and the flat metric. In [\gibras], the Euclideanised solution 
was interpreted as an instanton, and the analytic continuation to an E-brane type 
 solution was interpreted as an instability in which a wormhole expands at the speed of light, leading 
to a decay of the Minkowski-space vacuum. 

The E$4$-brane geometry   is 
$$
 ds^2=H^{-1/2}(dx_1^2+\dots+dx_4^2)+H^{1/2}
(-dt^2+dr^2+ r^2 d\www _4^2),
\eqn\efo$$
with 
$$   H=1+ {a^4\over (r^2-t^2) ^{2}}
\eqn\abc$$
The spatial slice $t=0$ is a wormhole with an infinite throat whose cross-section is an $S^4$, and 
which interpolates between flat space (at large $r$)
and the throat geometry $H^5\times S^4$ (at small $r$). 
As $t$ increases, the radius of the $S^4$ of fixed $r,x_1,...,x_4$ increases from $a$ at $t=0$ to infinity at $t=r$, and
 the wormhole expands at the speed of light.
For the complete solution with $r^2> t^2$, the area $r^2<t^2$ is excluded, and so as $t$ increases, an ever increasing region of the $\R^5$ parameterised by 
$r$  and the $S^4$ coordinates is excluded. Thus the wormhole 
  \lq eats up' the   whole of this $\R^5$, and such wormholes 
are spontaneously produced throughout $\R^5$ at a certain rate per unit 5-volume.
 This can  be interpreted as an instability of  the Minkowski space 
vacuum of the type $IIB^*$ theory, arising from the tunneling associated with   the D$4$-instanton.

In terms of the coordinates $\ss, \aa$ defined by \dssad, the solution for 
 the region $r^2> t^2$
is
$$
 ds^2=
\left({\ss^4 \over a^4+\ss^4}\right)^{1/2}
 dx_{\vert \vert} ^2  
 +\left({a^4+\ss^4 \over \ss^4}\right)^{1/2}
  d \sigma^2+
(a^4+\ss^4)^{1/2} d\hat  \www _5^2
\eqn\ebrspsd$$
where  $ d\hat \www _5^2$
is the 5-dimensional de Sitter metric of radius $1$.
As we have seen, this interpolates between flat space and $dS_5\times H^5$. 
A slice of constant $\ss$ is $\R^4\times dS_5$, where the radius of the de Sitter space is $(a^4+\ss^4)^{1/4}$
and increases from $a$ at the boundary $\ss=0$ to infinity as $\ss$ becomes large.
For fixed $\ss$ and $\aa$, the space is $\R^4\times S^4$, and as the de Sitter time $\aa$ evolves, the $S^4$ contracts to a
minimum radius $a$ at time $\aa=0$ and then expands.
An observer in the space-time   \ebrspsd\ would   interpret this as a universe expanding in four spatial dimensions, and the region $t^2>r^2$ in which 
$\ss$ is imaginary would not be  part of his universe. To him, the idea that there is a region outside his universe that is
being \lq eaten up' would not be testable.

Similarly, 
consider  the region $t^2 > r^2$.
In terms of the  coordinates $\tt,\bb$ defined in  \trtryru, the E$4$-brane metric is
$$
 ds^2=\left({\tt^4 \over a^4+\tt^4}\right)^{1/2}dx_{\vert \vert} ^2  
 -\left({a^4+\tt^4 \over \tt^4}\right)^{1/2} d \tau^2+
(a^4+\tt^4)^{1/2}  d\ti \www _5^2,
\eqn\abc$$
where
 $ d\ti \www _5^2$
is the metric on $H^5$ of \lq radius'  $1$. The space interpolates between flat space near  $\tt =  \infty$ or $\tt = 
-\infty$
and 
$dS_5 \times H^5$ near $\tt=0$.
A slice of constant time $\tt$ is $\R^4\times H_5$, where the radius of the hyperbolic space $H^5$ is $(a^4+\tt^4)^{1/4}$
and so starts being  infinitely large in the past at $\tt=-\infty$, decreases to radius $a$ at $\tt=0$ and then re-expands
indefinitely.
Again, an observer would see an expanding universe, and would not see anything outside his universe that was being \lq eaten
up' as his universe expands.

This suggests the following interpretation. For the type $II^*$ theories, the presence of ghosts leads to an instability of
the flat space vacuum. There is an instability associated with each of the D$q$-instantons for each $q$, which correspond to  worm-holes
expanding and eating up the transverse space, in the same way we have seen above for the D$4$ instanton. This is a \lq
tunneling to nothing', as in the decay of the Kaluza-Klein vacuum [\witkk]. However, whereas flat space is unstable, the
$IIB^*$ theory has a maximally supersymmetric $dS_5\times H^5$ solution, with a de Sitter cosmology. There are similar four
and seven dimensional de Sitter solutions of the 11-dimensional analogue of the type $IIA^*$ theory [\huku]. The  E$4$-brane
solution
\ebr\ leads to two separate geodesically complete solutions, each of which interpolates between the de Sitter cosmology and
flat space. The interpretation of the other E-branes will be considered further in [\huku].

However, it is also interesting to consider the full E$4$-brane solution \efo\ for all values of $r,t$.
The space-time is divided into two regions by the light-cone $r^2=t^2$, each of which is non-singular and complete, as
discussed above. As $t $ increases (with $t>0$), the exterior region $r^2>t^2$ gets smaller,  but now instead of a  \lq
tunneling to nothing', there is a tunelling to the interior solution $t^2>r^2$, which grows to fill up the transverse space,
and is flat in the limit $t\to \infty$. 
This could be interpreted as follows: there is tachyonic brane creating a spherical \lq shock-wave' in the
transverse space expanding at the speed of light. This divides the space into two regions, $t^2<r^2$ and $t^2>r^2$,
and distorts the geometry so that geodesics do not reach the boundary $r^2=t^2$, and communication between the two regions
is impossible.  Far away from the light-cone in either region, the space becomes flat. 
As the hole inside the expanding bubble has been filled in, this would represent a shock-wave, not an instability.
An observer in the exterior region $r^2>t^2$ cannot tell whether there is nothing outside his universe, so that
the Minkowski vacuum can be said to have decayed, or whether there is another region on \lq the far side',
separated from his observable universe by a shock-wave, in which case 
the Minkowski vacuum could be said to be regained once the shock-wave has passed.

\chapter{De Sitter   Topological Gravity and   Large $N$ Topological Field Theory}

The Euclidean \sym\ theory has ghosts but can be twisted to obtain a topological field theory, in which the physical states
are the BRST cohomology classes.
The B-model, for example, arises from twisting the $SO(4)$ Lorentz symmetry with an $SO(4)$ subgroup of the $SO(5,1)$
R-symmetry. However, the original theory has $SO(5,1)$ conformal symmetry, and the twisted theory should also be conformal.
This can be made manifest by twisting the $SO(5,1)$ conformal symmetry with the $SO(5,1)$
R-symmetry, so that there is a supercharge that is a singlet under the diagonal $SO(5,1)$. This gives a BRST charge of the
topological theory that is invariant under the twisted conformal group $SO(5,1)$.
The B-model obtained by twisting   $SO(4)$ is in fact conformally invariant under the diagonal  $SO(5,1)$ of the original
$SO(5,1)\times SO(5,1)$ symmetry.
The A model and the half-twisted model arise from twisting the $SU(2)_L$ subgroup of the $SO(4)\sim SU(2)_L\times SU(2)_R$
Lorentz group with the two different embeddings of $SU(2)$ in the R-symmetry $SO(5,1)$.

If the Euclidean \sym\ theory is dual to the type $IIB^*$ 
theory in $dS_5\times H^5$, then a   twisting 
of the \sym\ theory should correspond to a twisting of the
  type $IIB^*$  theory, which would then be
a topological string theory with a corresponding topological gravity limit.
This should be particularly useful for studying the large $N$ limit of the topological gauge theory, as the supergravity dual
should give an alternative means of calculating correlation functions which may in some cases be much easier than the direct
calculation.
The supersymmetry generators of the \sym\ theory correspond to the global supersymmetries of the supergravity theory, arising
from the 32 Killing spinors of the de Sitter vacuum. After the twisting, one  linear combination of  the \sym\ supersymmetry
generators becomes the BRST charge of the topological field theory, and so the corresponding  linear combination of  the
Killing supersymmetries should lead to the BRST transformation on the supergravity or superstring side. After twisting, the
Killing spinor corresponding to the BRST transformations becomes a \lq Killing scalar' or constant. Whereras only a special
class of spaces admit Killing spinors, all admit Killing scalars, and so the superstring theory might be expected to remain
topological on a more general class of background. These topological models will be discussed
further elsewhere. 
 
\ack
{I would like to thank Jos\' e Figueroa-O'Farrill, Jerome Gauntlett, Gary Gibbons and  Ramzi Khuri for helpful discussions.}

 \appendix 

{In the   analysis of section 10, use was made of the Euclideanised type II supergravity theories, and it may be 
useful to discuss these here. We will ignore the issues of the fermionic sector and concentrate on the bosonic actions.
Recall that for $D$-dimensional Einstein-Maxwell theory, the Wick rotation is accompanied by
the analytic continuation $A_0 \to -iA_0$ of the time component of the potential, so that the connection one-form remains real
and the resulting Maxwell action is positive definite and 
$SO(D)$ invariant; see e.g. [\haweuc]. The Euclideanised Lagrangian is then proportional to $-R+F^2$.
Then solutions (e.g. Euclideanised Reissner-Nordstrom) carrying  only a magnetic   charge are real, 
while those carrying electric 
charge must have an imaginary electric charge, so that $A_0$ is pure imaginary.  
If instead one were to continue the
spatial part of the potential  $A_{i} \to iA_i$ instead of the time component, the
resulting theory would have a Lagrangian   proportional to $-R-F^2$.
with a negative definite Maxwell action.
However, with this continuation, the  Wick rotation of the electrically charged solutions would be real, 
but the magnetically charged ones would have an imaginary magnetic charge and imaginary vector potential.
This is reflected in the fact that  if the
 the vector field is dualised to a $D-3$ form gauge field $\ti A$, then
$F^2=- \ti F^2$ so that the sign of 
 the
kinetic term of the dual gauge field $\ti A$ would be the opposite to that   of $A$.
In the Euclidean path integral, it is the action  $-R+F^2$ that is used if one 
integrates over $A$ (as opposed to $\ti A$) as it is positive (apart from the conformal gravitational mode,
which can be separately continued), while if one integrates over $\ti A$, one uses $-R+\ti F^2$.

The Euclideanised bosonic part of the IIA supergravity action is (see e.g. [\beck])
$$\eqalign{
I_{IIA}= 
\int
 d^{10} x &
\sqrt{g}\left[
 e^{-2 \Phi} \left(- R- 4(\partial   \Phi )^2  
+H^2 
\right)\right.
\cr &
\left. +G_2^2 +G_4^2 \right] + {4i\over \sqrt 3}\int G_4 \wedge G_4 \wedge B_2 + \dots
\cr}
\eqn\twoae$$
This results from the Wick rotation $ t \to it $ 
accompanied by the analytic continuation $A_{0i....j} \to -i A_{0i....j}$ for the electric components of  the
 gauge fields $B_2, C_1,C_3$, so that the $n$-form potentials remain real and the Euclideanised action for these 
 fields is positive. The 
  Wick-rotation of the magnetically charged $p$-branes 
(i.e. those with $p \ge 4$) to $p+1$-instantons are real solutions of  the Euclideanised theory, while the electrically
charged ones (i.e. those with $p < 4$) rotate to solutions in which the electric brane charge is imaginary, so the solutions have a real metric  and dilaton, but the $p+1$ form gauge field is pure imaginary.  
On the other hand, similar steps lead to the following action for the Euclideanised version of the $IIA^*$ theory 
$$\eqalign{
I_{IIA^*}= 
\int
 d^{10} x &
\sqrt{g}\left[
 e^{-2 \Phi} \left(- R- 4(\partial   \Phi )^2  
+H^2 
\right)\right.
\cr &
\left. -G_2^2 -G_4^2 \right] - {4i\over \sqrt 3}\int G_4 \wedge G_4 \wedge B_2 + \dots
\cr}
\eqn\twoaes$$
Then the D$n$-instantons  from Wick rotating E$n$-branes with $n\le 4$ are real, while the those from Wick rotating E$n$-branes with $n> 4$ carry imaginary charge and couple to an imaginary RR potential. Moreover, the RR kinetic terms are negative definite (unless dualised).
However, instead of  analytically continuing the electric RR potentials $C_{0i..j}\to -i
C_{0i..j}$, one could instead continue the spatial components $C_{i..j}\to i
C_{i..j}$,
and this gives the   Euclideanised IIA  action \twoae.
If one is to integrate over the RR potentials $C_1,C_3$, then this is the Euclideanised action that is appropriate 
as the RR kinetic terms are positive, while if one were to integrate over the dual potentials $\ti C_7
, \ti C_5$, it would be the Euclideanised   action \twoaes, expressed in terms of   $\ti C_7
, \ti C_5$, that would be used.
This is the sense in which the IIA and the $IIA^*$ theories can be continued to the same 
Euclideanised action, which is \twoae\ if one integrates over $C_1,C_3$, but would have been
\twoaes\ if one had integrated over the dual potentials.

Similar considerations apply to the IIB theory. 
The Euclideanised bosonic 
action can be taken to be
$$
\eqalign{I_{IIB}= 
\int
 d^{10} x &
\sqrt{g}\left[
 e^{-2 \Phi} \left( -R-4(\partial   \Phi )^2  
+H^2 
\right)\right.
\cr &
\left. +G_1 ^2+ G_3 ^2+ G_5 ^2 \right] + \dots
\cr}
\eqn\twobe$$
but now the field equations cannot be supplemented by the constraint $G_5=*G_5$, as in 10 Euclidean dimensions $**G_5=-G_5$ and this would imply $G_5=0$, while $G_5=i*G_5$
would require a complex $C_4$. 
This is the Euclideanised  action for either the IIB or $IIB^*$ theories if one integrates over the 
potentials $C_0,C_2,C_4$ for either the
IIB or the $IIB^*$ theories, but the sign of   the kinetic terms would  
be reversed if one were to regard the dual potentials as fundamental.
With the signs of \twobe, the D0 and D2   instantons carry imaginary charge; in particular, the D0 instanton or D  instanton has an imaginary RR scalar background, but the action governing
the fluctuations in $C_0$ is positive and indeed the action \twobe\ has scalars taking values in $SL(2,\R)/SO(2)$.
However, 
the action obtained from this by $C_0 \to -iC_0$ would 
have a D-instanton with real $C_0$ and scalars taking values in $SL(2,\R)/SO(1,1)$, as in [\sevbrane].
This would be the appropriate action
if one were to integrate over the dual potential $\ti C_8$, instead of  $C_0$ as $d\ti C_8^2=-dC_0^2$. }

\refout
\bye